\documentclass[iop,twocolappendix]{emulateapj}
\pdfoutput=1
\usepackage[breaklinks,colorlinks,citecolor=blue]{hyperref}
\usepackage{hyperref}
\usepackage{pdfpages}
\usepackage[all]{hypcap}
\usepackage{xcolor}
\usepackage{verbatim,epstopdf,graphicx,bigints, amsmath}
\newcommand{\myemail}{bsalmon@physics.tamu.edu}

\newcommand{\Mstar}{\hbox{$\mathrm{M}_\star$}}

\newcommand{\eg}{e.g.,}
\newcommand{\ie}{i.e.,}
\newcommand{\lir}{$L_{\text{TIR}}$}
\newcommand{\Rv}{$R_{\text{V}}$}

\newcommand{\Lsun}{\hbox{L$_\odot$}}

\newcommand{\ebv}{\hbox{${E(B-V)}$}}
\newcommand{\hst}{\textit{HST}}
\newcommand{\spitzer}{\textit{Spitzer}}
\newcommand{\herschel}{\textit{Herschel}}

\def\ergcm2s{\ifmmode {\rm\,erg\,cm^{-2}\,s^{-1}}\else
                ${\rm\,ergs\,cm^{-2}\,s^{-1}}$\fi}

\newcommand{\bdash}[1]{{\operatorname{#1}}}

\def\Snospace~{\S{}}

\newcommand{\zed}{$z$}
\hypersetup{
    colorlinks=true,
    linkcolor=blue,
    filecolor=magenta,      
    urlcolor=cyan,
}
\definecolor{venetianred}{rgb}{0.78, 0.03, 0.08}
\newcommand{\referee}{}

\shorttitle{Breaking the Curve with CANDELS}
\shortauthors{Salmon et al.}

\begin{document}
\title{Breaking the Curve with CANDELS: A Bayesian Approach to
Reveal the Non-Universality of the Dust-Attenuation Law at High Redshift}
\author{Brett Salmon$^{1,\dagger}$, Casey Papovich$^{1}$, James Long$^{2}$, S. P. Willner$^{3}$,
Steven L. Finkelstein$^{4}$,
Henry C. Ferguson$^{5}$,
Mark Dickinson$^{6}$,
Kenneth Duncan$^{7,8}$,
S. M. Faber$^{9}$,
Nimish Hathi$^{10}$,
Anton Koekemoer$^{5}$,
Peter Kurczynski$^{11}$,
Jeffery Newman$^{12}$, 
Camilla Pacifici$^{13}$,
Pablo G. P\'erez-Gonz\'alez$^{14}$,
Janine Pforr$^{10}$\\
}
\affil{
$^{1}$George P. and Cynthia W. Mitchell Institute for Fundamental Physics and Astronomy,\\
Department of Physics and Astronomy Texas A\&M University, College Station, TX 77843, USA, \\
$^{2}$Department of Statistics, Texas A\&M University, College Station, TX 77843-3143, USA, \\
$^{3}$Harvard-Smithsonian Center for Astrophysics, Cambridge, MA 02138, \\
$^{4}$Department of Astronomy, The University of Texas at Austin, Austin, TX 78712, USA, \\
$^{5}$Space Telescope Science Institute, Baltimore, MD, USA, \\
$^{6}$National Optical Astronomy Observatories, Tucson, AZ, USA, \\
$^{7}$University of Nottingham, School of Physics \& Astronomy, Nottingham NG7 2RD, \\
$^{8}$Leiden Observatory, Leiden University, NL-2300 RA Leiden, Netherlands, \\
$^{9}$UCO/Lick Observatory, Department of Astronomy and Astrophysics, University of California, Santa Cruz, CA 95064, USA, \\
$^{10}$Aix Marseille Universit\'{e}, CNRS, LAM (Laboratoire d`Astrophysique de Marseille) UMR 7326, 13388, Marseille, France
$^{11}$Department of Physics and Astronomy, Rutgers, The State University of New Jersey, Piscataway, NJ 08854, USA, \\
$^{12}$Department of Physics and Astronomy, University of Pittsburgh and PITT-PACC, 3941 OHara St., Pittsburgh, PA 15260, USA, \\
$^{13}$Astrophysics Science Division, Goddard Space Flight Center, Code 665, Greenbelt, MD 20771, USA, \\
$^{14}$Departamento de Astrof\'{\i}sica, Facultad de CC.  F\'{\i}sicas, Universidad Complutense de Madrid, E-28040 Madrid, Spain
}
\altaffiltext{}{$\dagger\ $\myemail}
\submitted{Resubmitted to ApJ on 5/4/16}

\begin{abstract} 
Dust attenuation affects nearly all observational aspects  of galaxy
evolution, yet very little is known about the form of the
dust-attenuation law in the distant Universe.  Here, we model the
spectral energy distributions (SEDs) of galaxies at $z\sim$~1.5--3
from CANDELS with rest-frame UV to near-IR imaging under different
assumptions about the dust law, and compare the amount of inferred
attenuated light with the observed infrared (IR) luminosities.  Some
individual galaxies show strong Bayesian evidence in preference of one
dust law over another, and this preference agrees with their observed
location on the plane of infrared excess ($IRX$, \lir/$L_{\text{UV}}$)
and UV slope ($\beta$).    We generalize the shape of the dust law
with an empirical model, $A_{\lambda,\delta}=\ebv\ k_\lambda\
(\lambda/\lambda_V)^\delta$ where $k_\lambda$ is the dust law of
\cite{Calzetti00}, and show that there exists a correlation between
the color excess \ebv\ and tilt $\delta$ with
${\delta=(0.62\pm0.05)\log(E(B-V))+}$(0.26~$\pm$~0.02).  Galaxies with
high color excess have a shallower, starburst-like law, and those with
low color excess have a steeper, SMC-like law.  Surprisingly, the
galaxies in our sample show no correlation between the shape of the
dust law and stellar mass, star-formation rate, or $\beta$.   The
change in the dust law with color excess is consistent with a model
where attenuation is caused by by scattering, a mixed star-dust
geometry, and/or trends with stellar population age, metallicity, and
dust grain size.  This rest-frame UV-to-near-IR method shows potential to
constrain the dust law at even higher ($z>3$) redshifts. 
\end{abstract}

\section{Introduction} 
Our knowledge of star-formation rates (SFRs) among the majority of 
\referee{${z > 4}$ galaxies is, except in rare cases,} limited to
observations in the rest-frame ultraviolet (UV) where the effects of the dust
attenuation are most severe and lead to large systematics.
\referee{Galaxy surveys at the highest redshifts} are predominantly limited to studying the
rest-frame ultraviolet (UV)-to-near infrared (NIR) spectral energy
distribution (SED). The dust attenuation at this critical portion of
the SED cannot be dismissed even at $z= 7-8$, considering the mounting
observations of high-redshift dusty star-forming galaxies,
sub-millimeter galaxies, and quasars \citep{Wang08,
Casey14b, Mancuso16}. In addition, while there is no shortage of
observations \referee{and} simulations that offer potential mechanisms for dust
production in the early universe \citep{Todini01,
Gall11, Gall11b, Gall11c, Ventura14}, it is still uncertain how, and to
what degree, these mechanisms influence the wavelength-dependence of
attenuation at high redshift. 

The nuances of dust geometry, extinction, and scattering from the
interstellar medium (ISM) and star-forming regions are often
conveniently packaged into a ``recipe" of reddening
\citep{Calzetti97b}, parameterized by a wavelength-dependent curve of
the total-to-selective extinction \citep[][and references
therein]{Witt00},\vspace{-0.2cm}
\begin{equation} \label{equ:klambda}
k_{\lambda} = A_\lambda/E(B-V)\ 
\mathrm{and}\ R_{\text{V}} = A_\text{V}/E(B-V)\ ,
\end{equation} 
where $A_\lambda$ is the total extinction in magnitudes at wavelength
$\lambda$ and \ebv\ is the color excess of selective extinction.
We emphasize the distinction that dust ``extinction" accounts for the
absorption and scattering of light out of the line of sight, whereas
``attenuation" also accounts for the spatial scattering of light into
the line of sight for extended sources such as galaxies. We refer to
both extinction and attenuation models as ``dust laws" for brevity.
Successful empirical and analytic dust laws have been used for decades
as a necessary \emph{a priori} assumption when inferring fundamental
physical properties of \referee{distant galaxies \citep{Meurer99,
Papovich01}}.

Dust laws are already known to be non-universal across all galaxy
types from derivations of the Small and Large Magellanic Cloud (SMC
and LMC) and Milky Way dust laws, as well as dust attenuation in
$z$$<$1 galaxies \citep{Conroy10}.  For example, \cite{Kriek13} have
shown that the form of the dust law can vary significantly at $z<$2 as
a function of galaxy type, and in some cases it differs strongly from
the conventionally assumed \cite{Calzetti00} prescription, derived
from local UV-luminous starbursts.  The conditions that produce these
unique dust laws are complex. They depend on the covering factor,
\referee{the dust grain size (which is dependent on the observed 
composition and metallicity), and line-of-sight
geometry}  and can therefore change when galaxies are viewed at
different orientations \citep{Witt00, Chevallard13} or stellar
population ages \citep{Charlot00}. 

\referee{Changes to the observed star-dust geometry, that is,} the relative geometry
between stars and dust grains, produce different dust laws 
even for galaxies of a similar type. For example, observations of the
infrared excess ($IRX$ $\equiv\ L_{\mathrm{TIR}}/L_{\mathrm{UV}}$) and
the UV slope \referee{\citep[$\beta$, where $f_\lambda \propto \lambda^\beta$ over 
1268$<\lambda<$2580 \AA][]{Calzetti94}} have shown
that star-forming galaxies bracket a range of attenuation types from
starburst to SMC-like attenuations \citep{Buat11, Buat12,
Munoz-Mateos09, Overzier11}. The position of galaxies on the
$IRX-\beta$ plane suggests that a single dust-attenuation prescription
is incapable of explaining all observations \citep{Burgarella05,
Seibert05, Papovich06, Boquien09,Casey14}. 

Although star-forming galaxies have a variety of attenuation
scenarios, it is possible to infer their dust geometries by
correlating their inferred dust laws with physical properties. 
For example, \cite{Reddy15}
studied a sample of $z\sim2$ galaxies and found that the differences
in attenuation between gas and stars are correlated with the galaxy's
observed specific SFR (sSFR $\equiv$ SFR/\Mstar), potentially a byproduct of
the visibility of star-forming birth clouds.  If the dust law is
dependent on star-formation activity, then it may be different at
earlier epochs ($z>2$). \referee{This follows intuition because the intensity of
star-formation and ionization conditions, which directly influence the
attenuation conditions, have been shown to evolve with redshift}
\citep{Madau14,Steidel14,Casey14,Shimakawa15, Shapley15, Sanders15}.
These attenuation conditions are regulated by the formation, destruction, and
spatial distribution of dust grains, and this cycle is one of the most
poorly quantified processes in galaxies.  \referee{One reason to seek
evidence for the dust law is to place constraints on the observed 
dust grain size, which can be used to infer limits on dust production 
by SNe and AGB stars, especially given the maximum stellar population
ages at the redshifts of distant galaxies.} 

\referee{While the dust law gives clues to the underlying grain size
distribution in a broad sense \citep{Gordon00}, it
is difficult to connect the current grain sizes to their production
sources due to their complex history of growth, destruction,
and recycling over short timescales \citep{JonesAP13}.  In addition,
the dust production sources themselves, such as supernovae (SNe),
asymptotic giant branch (AGB) stars, or Population III stars have
changed in relative strength over cosmic timescales \citep{Morgan03,
Nozawa03}.  Contraints on the dust law can be used
to infer the observed dust grain sizes, which is
helpful when modeling the evolution of dust grain production and
evolution \citep[e.g., in high-redshift quasars][]{Nozawa15}.  A
better understanding of these mechanisms would help to constrain metal
buildup and galactic feedback \citep{Gall11, Dave11b}.}

\referee{In addition,} both the scale and the shape of the dust law
affect the interpretation of galaxy SFRs, the evolution of the SFR
density, and the evolution of the intergalactic medium (IGM) opacity.
For example, \cite{Smit14} showed that the measured $z\sim 7$ sSFR
changes by nearly an order of magnitude depending on the assumed
prescription of dust attenuation.  It is clear that new methods must
be developed to determine the shape of the dust law in the distant
universe.



Our goal in this work is to provide evidence for the dust law at 
high redshifts using the
information from galaxies' rest-frame UV-to-NIR SEDs.  We use a
Bayesian formalism that marginalizes over stellar population
parameters from models of the galaxy SEDs \citep{Salmon15}.   This
allows us to measure evidence in favor of one dust law over another
for individual galaxies.  We show that the favored dust
laws are consistent with the galaxies' locations on the
$IRX-\beta$ diagram for a sample of galaxies at $1.5 < z < 3.0$ with
mid-IR imaging, where we can verify that the predicted attenuation
agrees with the $IRX$. 

This work is organized as follows. \autoref{sec:DataSample} outlines
our photometric and IR data, redshifts, and sample selection, as well as our calculations
of IR luminosities and $\beta$. \autoref{sec:SEDfitting} describes the
framework of our SED-fitting procedure, including the stellar
population models and dust laws. \autoref{sec:BayesFactor} defines the
use of Bayes factors as our selection method, and
\autoref{sec:DeltaMethod} defines our parameterization of the
dust law. \autoref{sec:Results} shows the main results of the paper,
where we use our Bayesian technique to quantify the evidence that star-forming
galaxies at $z$$\sim$1--2 have a given dust law,
 using CANDELS \emph{Hubble Space Telescope} (\hst) and \spitzer\ 
data spanning the rest-frame UV-to-NIR SED. 
We then show that the UV color and thermal
IR emission (measured from mid-IR data) of these galaxies match the
properties of their predicted dust law. 
\autoref{sec:Discussion} discusses the implications and physical
origins of our results, as well as comparisons to previous work and
dust theory. Finally, \autoref{sec:Conclusions} summarizes our main
conclusions. We assume concordance cosmology such that $H_0$ = 70 km s$^{-1}$ Mpc$^{-1}$,
$\Omega_{\text{M,0}}$ = 0.3 and $\Omega_{\Lambda,0}$ = 0.7. 


\section{Data, Redshifts, and Sample Selection}\label{sec:DataSample}
\subsection{Photometry: CANDELS GOODS Multi-wavelength Data}\label{sec:candelsData}
This work takes advantage of the multi-wavelength
photometry from the GOODS North and South Fields \citep{Giavalisco04}, 
the CANDELS survey \citep{Grogin11, Koekemoer11},
the WFC3 Early Release Science program \citep[ERS][]{Windhorst11},
and the Hubble Ultra Deep Field \citep[HUDF][]{Beckwith06,Ellis13,
Koekemoer13, Illingworth13}.
We define magnitudes measured by \emph{HST} passbands with the
ACS F435W, F606W, F775W, F814W and F850LP as $B_{435}$, $V_{606}$,
$i_{775}$, $I_{814}$, and $z_{850}$  and with the WFC3 F098M, F105W,
F125W,  F140W, and F160W as $Y_{098}$, $Y_{105}$, $J_{125}$,
$JH_{140}$, and  $H_{160}$, respectively.   Similarly, bandpasses
acquired from ground-based observations include the VLT/ISAAC $K_s$
and VLT/HAWK-I $K_s$ bands. We refer to \cite{Guo13} for more details on the
GOODS-S dataset, and Barro et al. (in prep.) for the GOODS-N dataset.

As applied by \cite{Salmon15}, we include an additional
uncertainty,  defined to be 10\% of the flux density per passband of
each object. This accounts for any systematic uncertainty such as
flat-field variations, PSF and aperture mismatching, and local
background subtraction. The exact value was chosen from series of
recovery tests to semi-analytic models applied by \cite{Salmon15}.
Including this additional uncertainty also helps to avoid situations
where a given model SED band serendipitously finds a perfect match to
an observed low-uncertainty band, creating a biased posterior around
local maxima.  The additional uncertainty was added in quadrature to
the measured uncertainties.

\subsection{IR Photometry: Spitzer and Herschel} \label{sec:IRdata}
We utilize imaging in the IRAC 3.6 and 4.5~\micron\ bands from the
\spitzer\ Extended Deep Survey \citep{Ashby13} to measure the
rest-frame NIR of the galaxy SED.  As described by \cite{Guo11},
the IRAC catalog uses the \hst\ WFC3 high-resolution imaging as a
template and matches to the lower-resolution images using TFIT
\citep{Laidler07} to measure the photometry. 

In order to verify the dust-attenuation law derived from the rest
UV-to-NIR data, we require a measure of the rest UV-to-optical light
reprocessed by dust and reemitted in the far-IR.  Conventionally, the
important quantities are the ratio of the observed IR-to-UV
luminosities, $L(IR)/L(UV)$, which measures the amount of reprocessed
light,  and the UV-spectral slope, $\beta$, which measures the shape
of the dust-attenuation curve \citep[e.g.,][]{Meurer99, Charlot00,
Gordon00, Noll09, Reddy10}.  We used MIPS 24~\micron\ measurements from
the GOODS-\herschel\ program \citep{Elbaz11}, where the GOODS
IRAC~3.6~\micron\ data were used as prior positions to determine the
MIPS 24~\micron\ source positions. Then, PSF-fitting source extraction
was performed to obtain 24~\micron\ fluxes, which we require to be
$>3\sigma$ detections for our sample.  While we also examined galaxies
with \herschel\ PACS and SPIRE 100 to 250~\micron\ photometry, these
data were ultimately not included because they had no effect on the
results (see \autoref{sec:AppendixLIR}).


%
\begin{figure}        
\centerline{\includegraphics[scale=0.45]{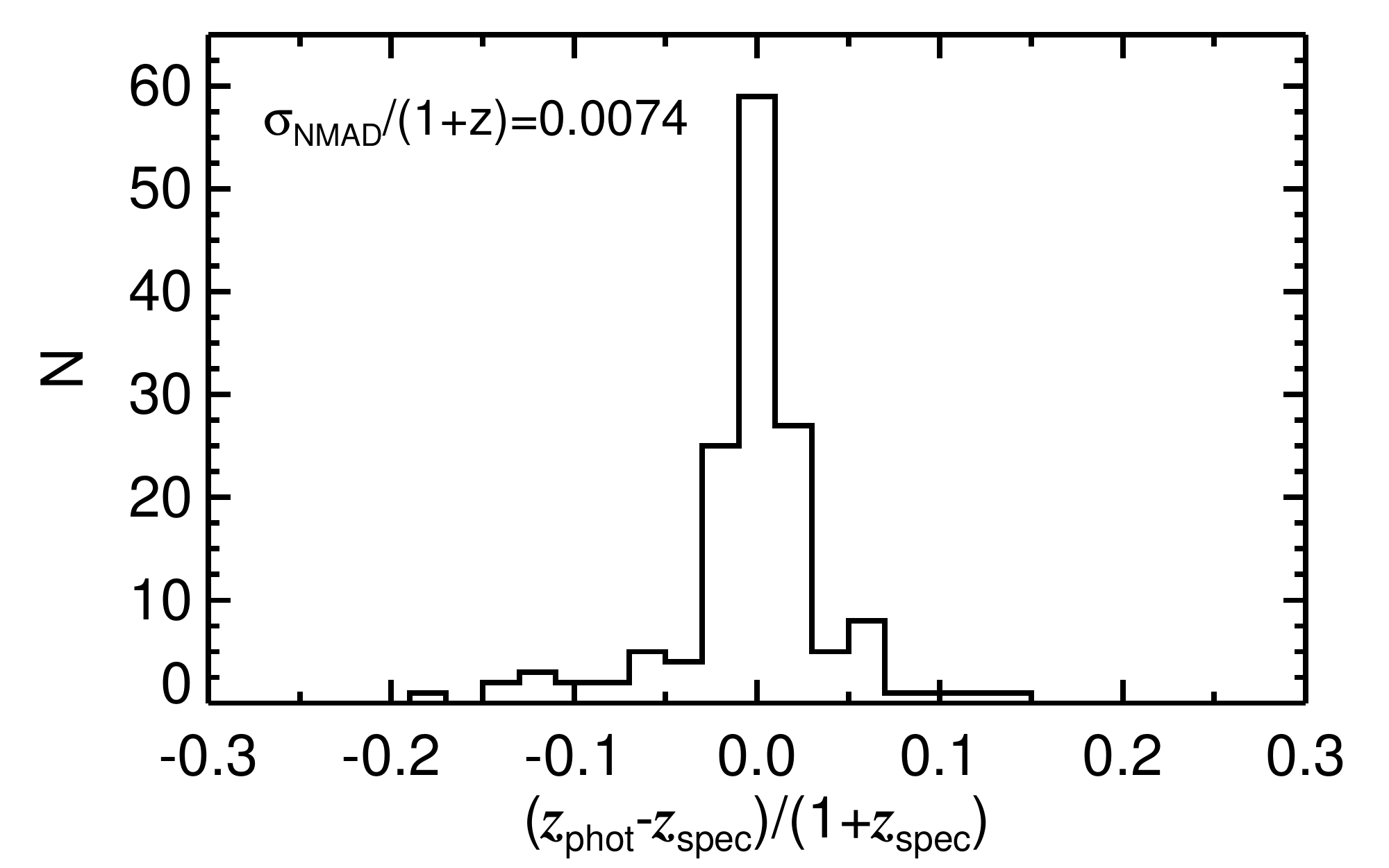}} 
\caption{
\referee{
The photometric redshift accuracy for galaxies that are in both the 
phot-$z$ and spec-$z$ samples. The $\sigma_\text{NMAD}$ gives the
68\% scatter of the distribution. 
}
}
\vspace{0.2 cm}
\label{fig:Redshifts}
\end{figure}
\setcounter{figure}{1}

\subsection{Redshifts} \label{sec:redshifts}
To minimize uncertainties in SED-fitting owing to redshift errors, we
selected objects that have the highest quality spectroscopic redshifts.  The
spectroscopic redshifts are a compilation (Nimish Hathi \&\ Mark
Dickinson, private communication) from several published and
unpublished studies of galaxies in GOODS-S \citep[][Weiner et al. (unpublished)]{Mignoli05,
Vanzella08, Balestra10, Popesso09, Doherty05, Kriek08, Fadda10} and
GOODS-N \citep[][]{Reddy06, Daddi09}. 
We define the sample of galaxies with high-quality redshifts as the
``spec-$z$" sample, but later we consider the full sample with
photometric redshifts, which we call the ``phot-$z$" sample.

The primary goal of this work is to determine the ubiquity of the
dust-attenuation law at the peak of cosmic SFR density. When deriving
properties of distant galaxies, we must naturally consider how our
results are dependent on the assumed redshift of each galaxy. This can
be done in two ways. First, we explore how our results depend on
redshift accuracy by testing how our results vary if we use
photometric redshifts for galaxies rather than their spectroscopic
redshifts.
Second, we determine how the results of the spec-$z$ sample differ
from a larger sample of galaxies with photometric redshifts. The
former test addresses how photometric redshift accuracy in general affects the
methods and results, while the latter test addresses if the photometric redshift
accuracy within a larger sample is sufficient to reproduce the
spectroscopic-redshift results. In addition, a photometric-redshift sample can reveal biases in
the spec-$z$ sample because the latter is likely biased towards the
brighter, bluer galaxies.

We used photometric redshifts that were derived following the methods
by \cite{Dahlen13}, who developed a hierarchal Bayesian technique to
convolve the efforts of eleven photometric redshift investigators in
the CANDELS team. The photometric-redshift estimates of GOODS-S are
taken from \cite{Santini15} and those of GOODS-N are taken from Dahlen
et al. (2015 in prep.). The GOODS-N photometric-redshift estimates
also take advantage of SHARDS-grism narrow-band data.  We take the
photometric redshift as the median from the combined full $P(z)$
distributions of nine GOODS-N and six GOODS-S photometric-redshift
investigators.  

\referee{\autoref{fig:Redshifts} shows the accuracy of the photometric
redshifts when compared to galaxies with known spectroscopic
redshifts. We estimate the photometric-redshift accuracy from the
normalized median absolute deviation \citep{Brammer08} which gives a
68\% scatter of the distribution of $\sigma_{\text{NMAD}}/(1+z) =
0.0074$.  In addition, 94\%\ of the sample has a quality of
${|z_\text{phot}-z_\text{spec}|/(1+z_\text{spec})<0.1}$.  This gives
us confidence that the redshifts of the phot-$z$ sample are well
determined. }


%
\begin{figure}        
\centerline{\includegraphics[scale=0.4]{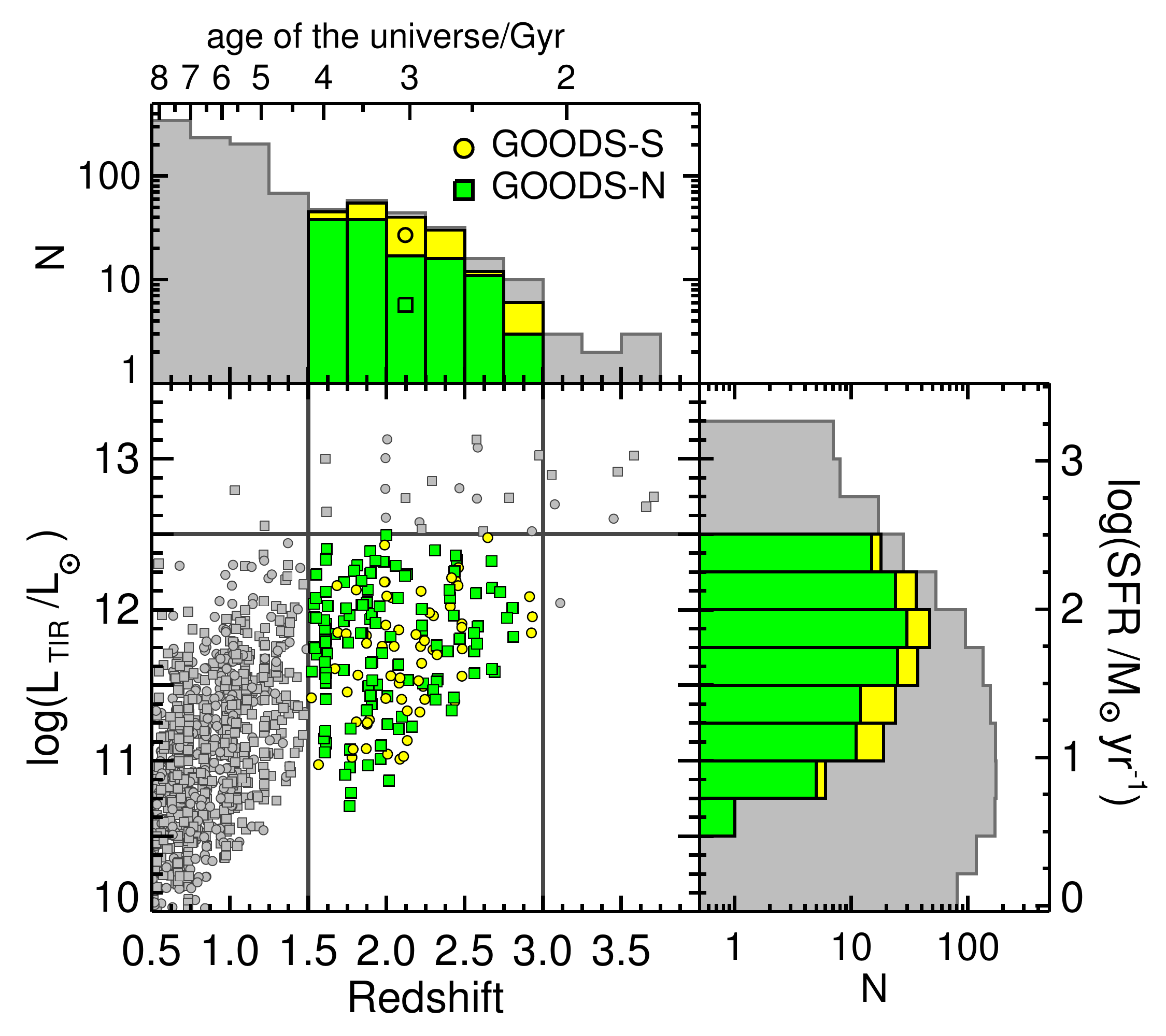}} 
\caption{The log of total IR Luminosity (\lir), which was determined
using a redshift-dependent conversion from $L_{24~\micron}$,
as a function of redshift.
Galaxies in our sample lie in GOODS-N (yellow) and GOODS-S (green) and
were restricted to $\log$ \lir $<12.5$ and
\referee{$1.5<z_{\text{spec}}<3.0$}. The
adjacent histograms compare the logarithmic distributions of our
sample to the parent sample. For reference, the top and right axes
show the age of the universe and the SFR respectively. 
}
\vspace{0.2 cm}
\label{fig:lir}
\end{figure}
\setcounter{figure}{2}
\subsection{Sample Selection} \label{sec:sample}
We limited the sample to $z>1.5$, such that the ACS $B_{435}$ band
still samples the rest-frame far-UV (FUV, $\sim$1500~\AA), which is a
crucial portion of the SED when distinguishing between dust laws.
\autoref{sec:speczSEDresults} discusses the consequences of a galaxy
not having a band close to the FUV, due to the redshift or available
photometry.  We also required a $z < 3$ limit because the IR-selection of
sources at higher redshift correspond to objects with very bright IR
luminosities ({$\log$ \lir/\Lsun $>$ 12.5}), where the frequency of
objects dominated by AGN emission increases to $\sim60\%$
\citep{Nardini10}.  In addition, the upper redshift limit was chosen to
avoid significant redshift evolution within the sample.  

\referee{
Applying the redshift range of $1.5<z<3.0$ and requiring 24~\micron\ 
detections ($S/N >3$)} to the sample produces
an initial sample of 65 (554) GOODS-N and 123 (552) GOODS-S
spec-$z$ (phot-$z$) selected galaxies. A small number ($<5\%$ of the
spec-$z$ sample and $<2\%$ of the phot-$z$) of objects were identified
on or near bright stars and diffraction spikes, as well as at the
edges of the image \citep{Guo13} and were removed from all samples.  

We further identified galaxies that imply the presence of an active
galactic nucleus (AGN) from their IR or radio data \citep{Padovani11,
Donley12} or if they have known X-ray detections \citep{Xue11}. This
selection removes 6 (52) GOODS-N and 31 (108) GOODS-S sources in the
spectroscopic (phot-$z$) spec-$z$ sample.  Our final sample contains
56 (485) GOODS-N and 88 (432) GOODS-S galaxies in the fiducial
spec-$z$ (phot-$z$) sample.

\subsection{Calculation of Total Infrared Luminosities}
\label{sec:lir} One method to calculate the total infrared luminosity
(\lir) involves fitting broadband flux densities to a suite of look-up
tables that were derived from templates of local IR luminous galaxies
\citep{Elbaz11, Dale01,Dale02, Rieke09}. However, recent work has
shown that template-fitting can overestimate \lir\, especially when
the observed bands do not well sample the dusty SED
\citep[see][]{Papovich07,Overzier11}. At the redshifts of our sample,
46\% of our galaxies lack detections redward of 24~\micron\ (\ie\
\herschel\ PACS or SPIRE).  

\referee{Detailed studies have shown that the rest-frame mid-IR emission is an
excellent estimator for \lir\ for both local and high-redshift ($z<2.8$) 
galaxies once it has been properly calibrated \citep[][R13 hereafter]{Wuyts08, Rujopakarn13}.  
This conversion was developed using the fact that the average IR SEDs
of galaxies are governed by their IR surface densities
\citep{Rujopakarn11}, allowing bolometric corrections to account for
high-redshift polyaromatic hydrocarbon (PAH) emission. 
For the galaxies in our range of redshift and luminosity, R13 showed
that the scatter in \lir\ derived from the 24~\micron\ emission is very tight,
only 0.06 dex (see R13, their Figure~2). Therefore, we adopted the relation
from R13 (their equation~3) to derive \lir\ for the galaxies in our study 
using their observed 24~\micron\ emission and redshifts. }


\referee{The adopted} 24~\micron\ conversion was developed under
several relevant assumptions: it applies to $z\sim2$ galaxies
that lie on the SFR-stellar mass main \referee{sequence (most galaxies
in our sample are on the main sequence),} the galaxies are not
hyperluminous (\lir\ $< 10^{13} L_\Sun$), and the $\log$ \lir surface density
scales linearly with $\log$ \lir. These assumptions become important
for compact starburst galaxies and ULIRGs (\lir/\Lsun $>10^{12}$).
Nevertheless, these objects are rare, and fewer than \referee{25\%} of
galaxies have \lir/\Lsun $>10^{12}$ in both the phot- and spec-$z$
samples.  This fraction of the sample are not the galaxies that drive
the results of this work.

As a further check, 54\% of the galaxies in our spec-$z$ sample have
\herschel\ PACS and/or SPIRE data. 
The comparison using this data to calculate \lir\
can be found in \autoref{sec:AppendixLIR}, but in
short, the results of this work are unaffected by using fits to
\herschel\ data instead of the 24~\micron\ conversion to calculate
\lir.  

The distribution of \lir\ is shown in \autoref{fig:lir} as a
function of redshift for all 24~\micron-detected sources with
spectroscopic redshifts, including those within our redshift range.
\referee{For reference, we also show the SFRs corresponding to a given
\lir\ following conversions equation~8 of R13 which is similar to the
\cite{Kennicutt98} conversion with factors applied appropriate for a
\cite{Salpeter55} IMF.} This figure shows that galaxies in our sample
have IR luminosities ranging from $5 \times 10^{10}$ to \referee{$3
\times 10^{12}$ \Lsun, consistent with luminosities of LIRGs.}

\subsection{\referee{Calculation of the UV Luminosity}}\label{sec:UV}
\referee{
We derive the observed UV luminosity, $L_\mathrm{UV}$, that is, the UV
luminosity uncorrected for dust attenuation as follows. 
$L_{\mathrm{UV}}$ was determined by taking the average luminosity of
the best-fit SED in a 100~\AA\ bandwidth centered at 1500~\AA. There
is little dependence on the choice of SED models, such as the choice
of dust law or SFH, in determing $L_{\mathrm{UV}}$. Similar results
are also found when approximating $L_{\mathrm{UV}}$ as the luminosity 
of the observed band closest to rest-frame 1500~\AA. 
}

\subsection{Calculation of the UV Spectral Slope $\beta$} \label{sec:Beta}
The rest-frame UV slope is an important observational tool due to its
relative ease of measurement for the highest redshift galaxies
\citep[even to $z\sim10$, see][]{Wilkins15} and its sensitivity to
stellar population age, metallicity, and attenuation by dust.
Moreover, $\beta$ has often been used to estimate the dust
attenuation by extrapolating its well-known local correlation with
infrared excess \citep{Meurer95,Meurer99}. Studies of the origins 
of the scatter in the $IRX-\beta$ relation show that it depends on
metallicity, stellar population age, star-formation history, spatial
disassociation of UV and IR components, and the shape of the
underlying dust-attenuation curve, including the presence of the
2175~\AA\ absorption feature \citep{Gordon00, Buat05, Buat10, Reddy06, Munoz-Mateos09,
Boquien12}.  This raises concerns about generalizing the $IRX-\beta$
relation to higher redshifts \citep[\eg\ see the discussion by][]{Casey14}.

Historically, the methods used to calculate $\beta$ have been entirely
dependent on the available dataset. In the absence of UV continuum
spectroscopy \citep[the original method to determine
$\beta$,][]{Calzetti94}, we must calculate $\beta$ from the UV colors
provided by broadband photometry. Specifically, we calculated
$\beta$ from the best-fit SED following the methods of
\cite{Finkelstein12a}. We favor this method over a power-law fit to
the observed photometric bands for the following reasons. 

First, we ran simple tests \referee{on the stellar population models} to recover the
input $\beta$ with a power-law fit to
the bands with central wavelengths between rest-frame
$1200<\lambda<3000$~\AA.  The true $\beta$ is determined from stellar
population models by \cite{Kinney96}, using the spectral windows
defined by \cite{Calzetti94} after applying a range of \ebv.  This
method produced a systematic offset at all redshifts such that
{$\beta_{\text{true}}$ = $\beta_\text{phot} - 0.1$}, and at some
redshifts the recovery is off as much as $\Delta \beta=-0.5$.  
\bgroup
\def\arraystretch{1.5}
\begin{table*}[t!]
\label{tab:sedparms}
\caption{SED Fitting Parameters} 
\centering  
\begin{tabular}{|c  c  c  l |} 
\hline                         
Parameter & Quantity & Prior & Relevant Sections \\  
\hline                  
Redshift & fixed & spectroscopic redshifts, 1.5 $ \leq z_\text{spec}
\leq $ 3.0 & \S~\ref{sec:redshifts} \S~\ref{sec:speczResults} \\
         & fixed & photometric redshifts, 1.5 $ \leq z_\text{phot}
\leq $ 3.0 & \S~\ref{sec:redshifts} \S~\ref{sec:photzResults} \\
\hline
Age & 100 &  10 Myr to $t_\mathrm{max}$ 
              \footnote{The lower end of this range represents the
                minimum dynamical time of galaxies in our redshift
                range up to $t_\mathrm{max}$, which is the age of the
                Universe for the redshift of each object, which is up 
                to 4.2 Gyr at $z\sim 1.5$.}  & -- \\ 
\hline
Metallicity & 5 & $Z= 0.02, 0.2, 0.4, 1.0, 2.5~Z_\Sun$  & -- \\
\hline
\ebv\ \footnote{We fit to a range of color excess values, \ebv.
  This scales the dust-attenuation curve to achieve a wavelength-dependent 
  attenuation, {$A(\lambda) = k(\lambda) E(B-V)$}.}
  & 85 & \referee{Linear, $-0.6 < $$E(B$$-$$V) < 1.5$, 
$\Delta$$E($$B$$-$$V)=0.025$}  & \S~\ref{sec:DustCurves} \\
\hline
Attenuation prescription & fixed & starburst
\citep{Calzetti00} or SMC92 \citep{Pei92}  & \S~\ref{sec:DustCurves}
\S~\ref{sec:speczDeltaResults}\\
 & \referee{11}\footnote{\referee{This parameterized dust law is a power-law
deviation from the starburst dust law, similar to 
\cite{Noll09} (see \autoref{sec:DeltaMethod} for a detailed
definition).}} & \referee{$-0.6 < \delta < +0.4$, in steps of
$\Delta\delta=0.1$}
 & \S~\ref{sec:DeltaMethod} \S~\ref{sec:photzDeltaResults} \\
\hline
$f_{\text{esc}}$ & fixed & 0 & -- \\
 \hline
Star formation history
	      \footnote{The star formation history is defined as $\Psi(t)
= \Psi_0 \exp(t/\tau)$ such that a SFR that increases with cosmic
time has a positive $e$-folding time, $\tau$.   When the SFH is
allowed to vary as a fitted parameter, we consider rising and
declining histories (positive and negative $\tau$) seperately. The
long $e$-folding time of ${\tau = 100}$ Gyr is effectively a constant star-formation history.}
              & fixed & 100 Gyr (constant) &  \S~\ref{sec:SEDmodels} \\ 
              & \referee{10} & $\pm \tau=$ 0.1, 0.3, 1, 3, 10~Gyr &  \S~\ref{sec:SEDmodels} \\ 
\hline 
\end{tabular} 
\end{table*}
\egroup

Second, \cite{Finkelstein12a} saw a similar offset and scatter in
recovering $\beta$ from a single color or power-law fit. They promoted
calculating $\beta$ by using UV-to-optical photometry to find the
best-fit SED and using the UV spectral windows of \cite{Calzetti94} to
determine $\beta$. Their simulations reported a better recovery of
$\beta_{\text{true}}$ with no clear systematics and a scatter of
$\Delta \beta = \pm 0.1$ for galaxies at $z=4$. We therefore used the
best-fit model to calculate $\beta$, assuming a constant SFH and a
starburst \citep{Calzetti00} dust law \referee{(see
\autoref{sec:AppendixBeta} 
which shows that the results are not sensitive to changing the
derivation of $\beta$ to be a power-law fit to the photometry in the
rest-frame UV).}

One may be concerned that \referee{the adopted method makes $\beta$
sensitive to the assumed dust law of the SED models}.  However, the
best-fit SED will always provide a close match to the UV colors so
long as the assumed dust law does not have any extreme features such
as the excess of absorption at 2175~\AA\ or the almost broken
power-law rise in the far UV of the \cite{Pei92} extinction curve.  We
found similar results when calculating $\beta$ from the best-fit SED
when we allow the shape of the dust law to vary as a new parameter in
\autoref{sec:DeltaMethod}.

\section{Modeling Stellar Populations} \label{sec:SEDfitting}
The bulk of the methods and procedures of the SED fitting are described
by \cite{Salmon15}, which we summarize here including recent changes.
The SED fitting is Bayesian in nature, offering a mechanism to
determine the conditional probability for each desired physical
property of the galaxy.

\subsection{Bayesian Methods} \label{sec:SEDMethods}
Using Bayes' theorem, 
\begin{equation} \label{equ:Bayes} 
P(\Theta'|D) = P(D|\Theta')\ P(\Theta') / P(D) ,
\end{equation} 
we determine the posterior, $P(\Theta'|D)$, with parameters
$\Theta'=(\Theta\{ t_{\mathrm{age}}, E(B-V), Z \}, \Mstar$) and data,
$D$, under the \emph{a priori} probability of the parameters or
simply the ``prior", $P(\Theta')$. The likelihood, $P(D|\Theta')$, is
determined in the usual way using $\chi^2$ statistics \citep[\ie\
equation 2 of][]{Salmon15}.  The unconditional marginal likelihood of
the data, $P(D)$, often referred to as the Bayesian evidence\footnote{The
Bayesian evidence is occasionally denoted by $Z$.  We adopt the formal 
definition, $P(D)$, to avoid confusion with the
conventional astronomical symbol of metallicity.}, normalizes the
posterior such that the integrated posterior across all parameters is
equal to unity \citep{Jeffreys61,Heckerman95,Newton96}:
\begin{equation} \label{equ:BayesianEvidence} 
\mathrm{Bayesian~evidence}\equiv P(D) = \int_{\Theta} P(D|\Theta)\ P(\Theta)\  \mathrm{d}\Theta .
\end{equation} 
Calculating the unconditional marginal likelihood is a way to eliminate
the parameters $\Theta$ from the posterior (in \autoref{equ:Bayes})
through integration, leaving us with the probability of seeing the
data $D$ given all possible $\Theta$ \citep{Kass95}. The importance of
the marginal likelihood will be discussed further in
\autoref{sec:BayesFactor}.

Posteriors on individual parameters can be determined by marginalizing
over nuisance parameters. The strength of this Bayesian
approach is that the marginal probability of a given parameter is
conditional to the probability from the nuisance parameters. For
example, the posterior on \ebv\ is conditional
to the probability contribution from all stellar population ages,
metallicities, and star-formation histories. This approach 
is an alternative to using parameter results
taken from the best-fit (minimum $\chi^2$) model SED because it relies
on posterior integration instead of likelihood maximization. The
disadvantage of the latter is that small differences in $\chi^2$ or
an underrepresentation of measurement uncertainties can
result in best-fit models that are sporadic across the parameter
space, making results highly dependent on the SED template
assumptions \citep[see Figures 20 and 21 of][]{Salmon15}.  We therefore
favor using the median of each parameter's marginalized posterior over
results determined from the best-fit model, as supported by recent
literature \citep{Song15, Tanaka15, Smith15}. 

\subsection{Stellar Population Models} \label{sec:SEDmodels} 
Table~1 shows the ranges, \referee{quantity of values considered,} and priors of the SED fitting
parameters. Each combination of age, metallicity, and
\ebv\ produces an SED shape and associated $\chi^2$. The parameter
space \referee{was} constructed following the listed priors on each parameter.
We used \cite{Bruzual03} stellar population synthesis models with the
addition of nebular emission lines assuming an ionizing continuum
escape fraction of $f_\text{esc}$=0 \citep{Salmon15}. We assumed a
\cite{Salpeter55} initial mass function and \ion{H}{1} absorption
from line-of-sight IGM clouds according to \cite{Meiksin06}. 
The \cite{Meiksin06} IGM attenuation model includes higher order Lyman
transitions. Nevertheless, the assumption of IGM attenuation has minimal
effect on the results because few galaxies have photometry covering
wavelengths blueward of 1216~\AA.

The range of \ebv\ extends below zero for two reasons.  First,
consider the example where a Gaussian-shaped posterior for parameter $x$
peaks at $x=0$, but all probability at $x<0$ is set to zero. The $x$
corresponding to the median probability of such a posterior would be
biased to $x>0$, an artifact of the choice of parameter space.  This
was pointed out by \cite{Noll09}, who showed a bias to Bayesian
estimates of certain parameters, especially parameters such as \ebv\
whose posterior often peaks at the edge of the parameter space.
Second, negative values of \ebv\ are not necessarily unphysical.
There are some, albeit rare, situations where isotropic scattering by
dust in face-on galaxies can produce an enhancement of optical light
\citep[\ie\ $A_{\text{V}}<0$][]{Chevallard13}. 

We also considered how our results are dependent on the assumed shape of
the star-formation history. The star-formation history is known to be
a poorly constrained parameter in the fitting process
\citep[\eg][]{Papovich01,Noll09, Reddy12a, Buat12, Mitchell13}. While
it is not the motivation of this work to accurately fit the
star-formation history for individual galaxies, assuming a fixed
history may reduce flexibility in the parameter space and overstate
the perceived evidence between different dust laws. We therefore
considered three scenarios of the star-formation history (SFH):
constant, rising, and declining exponentially with cosmic time, with
ranges for the latter two cases described in Table~1. We take the
assumption of a constant history as our fiducial model, and we show in
\autoref{sec:AppendixSFH} that our main results are unchanged if we instead adopt
rising or declining star-formation histories. 

The stellar mass \referee{was} treated differently than the individual
parameters $\Theta$.  It is effectively a normalization of the SED,
given the mass-to-light ratio associated with the SED shape, hence the
distinction in \autoref{sec:SEDMethods} between $\Theta$, which
represents the parameters that actually drive the goodness of fit, and
$\Theta'$, which is those parameters and their associated stellar
mass. In this manner, the posterior in stellar mass was determined by
integrating the posterior rank-ordered by stellar mass to achieve a
cumulative probability distribution in stellar mass such that the
median is defined where the cumulative probability is equal to 50\%. 

\subsection{Known Dust Attenuation Curves} \label{sec:DustCurves}
\begin{figure}[!t] 
\centerline{\includegraphics[scale=0.5]{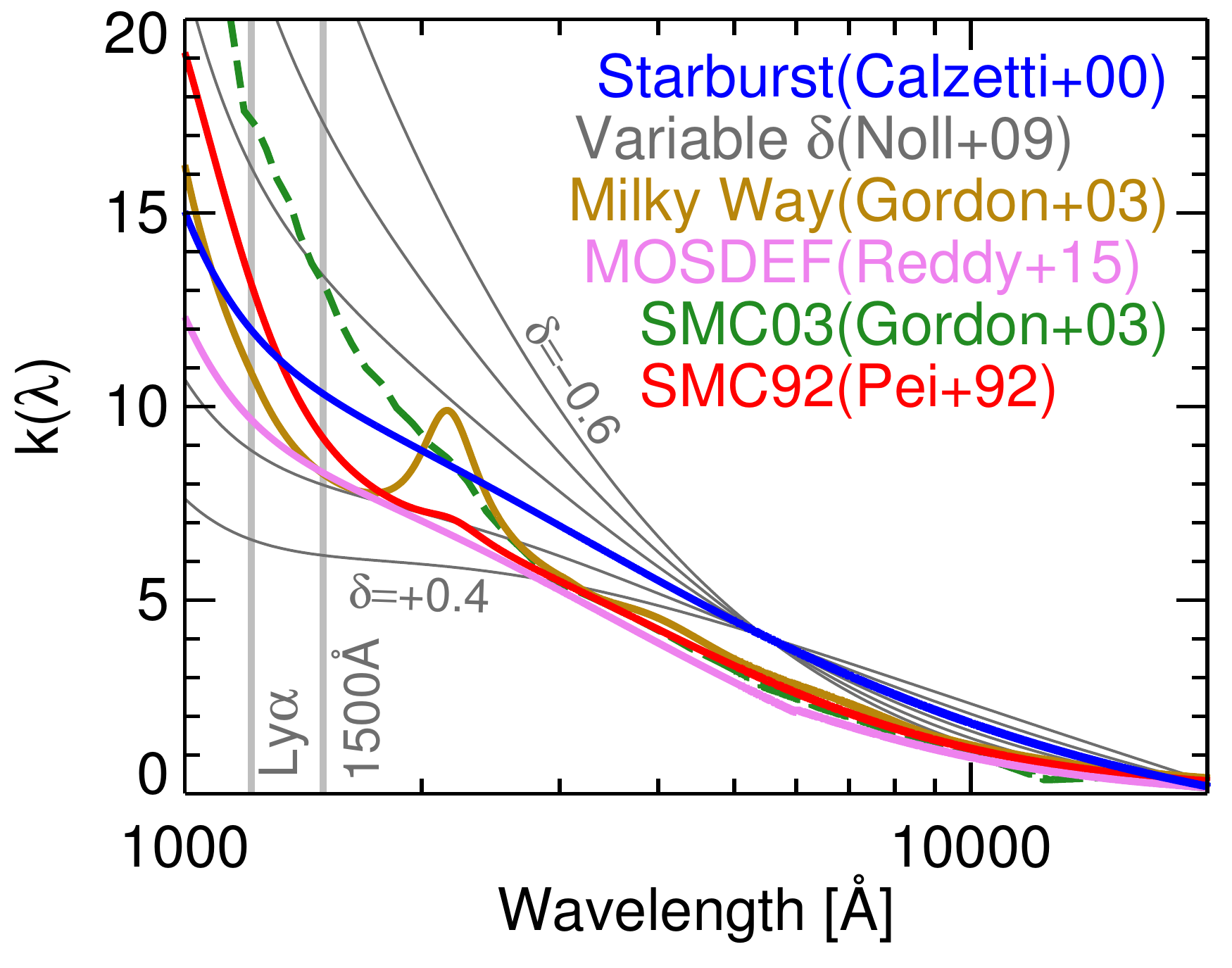}}
\caption{
A variety of common dust laws shown by their total-to-selective
extinction or attenuation as a function of wavelength.  The \cite{Pei92}
derivation of the SMC extinction (red, SMC92) will be used in this
work to compare to the starburst prescription derived by
\cite{Calzetti00} (blue). Other dust laws are also shown
including the MOSDEF (pink) attenuation curve derived from $z\sim2$
galaxies \citep{Reddy15}, and the Milky Way (gold) and SMC (SM03,
green, dashed) extinction curves derived by \cite{Gordon03}.  In
addition, we consider power-law deviations to the starburst curve
(\autoref{equ:delta}) to be more ($+\delta$) or less ($-\delta$) grey.
The wavelengths of 1500~\AA\ and the Lyman $\alpha$ emission line are
shown for reference.
} 
\label{fig:DustCurves} 
\end{figure}

The dust law was fixed during the fitting process (along with the
redshift, escape fraction, and star-formation history), although we 
individually considered a variety of commonly used dust laws. 
The curves of these dust laws are shown in \autoref{fig:DustCurves} and
include those of the empirically derived attenuation for local starburst
galaxies \citep{Calzetti00}, the Milky Way extinction \citep[which
showcases the strong 2175~\AA\ dust absorption feature,][]{Gordon03},
an empirically derived attenuation for $z\sim2$ star-forming galaxies
\citep[``MOSDEF",][]{Reddy15}, and two interpretations of the SMC
extinction: SMC92 \citep{Pei92} and SMC03 \citep{Gordon03}, hereafter.


%
\setcounter{figure}{3}

\begin{figure}[!t]
\centerline{\includegraphics[width=0.48\textwidth, trim=110 0 240 10,clip]{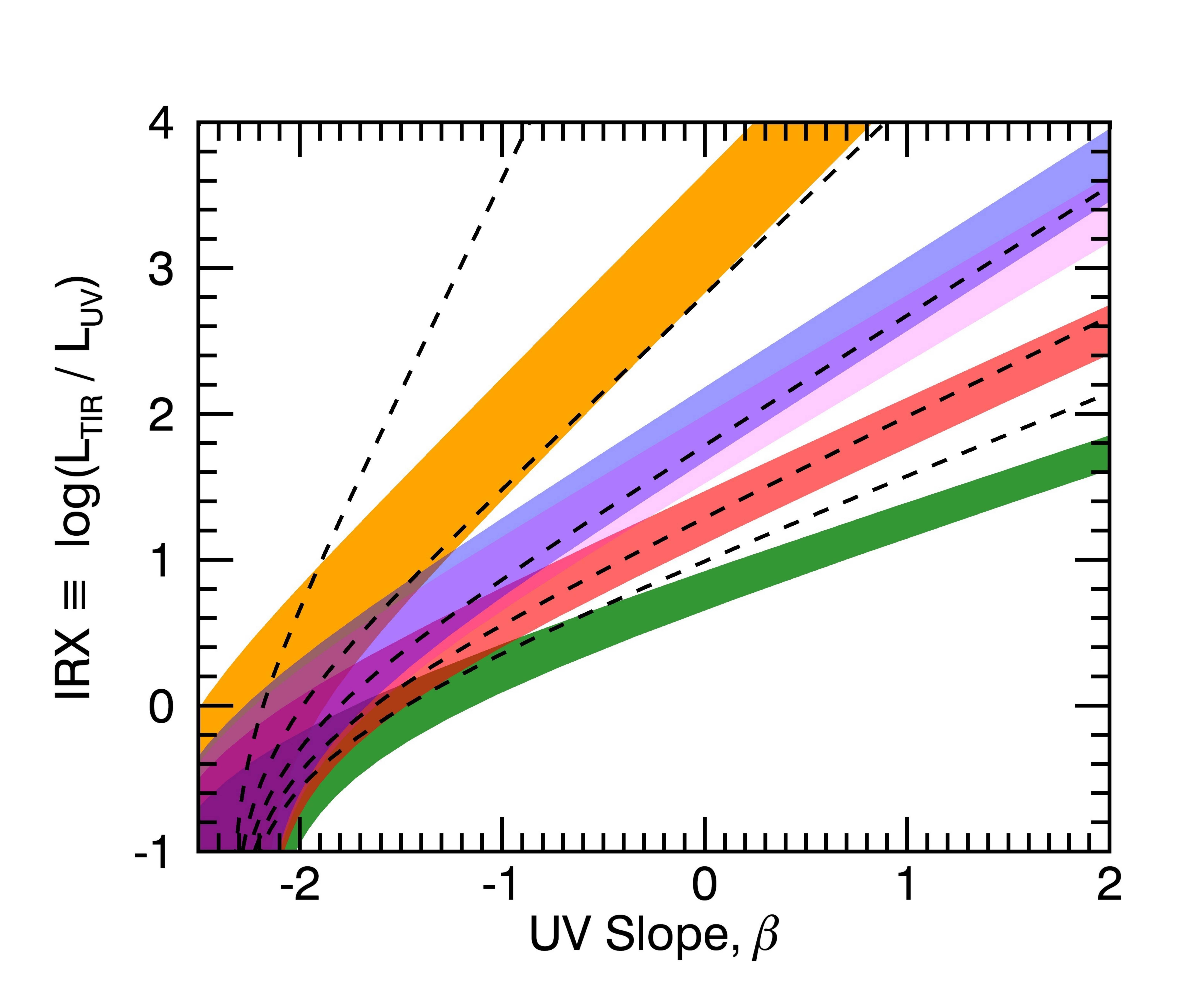}}
\caption{ 
The predicted locations of galaxies with different dust laws on the
plane of the UV slope $\beta$ and infrared excess
(\lir/$L_{\text{UV}}$). The colored swaths correspond to the same dust
laws as in \autoref{fig:DustCurves}, clockwise from top left: Milky
Way, starburst, MOSDEF, SMC92, and SMC03. The width of each
$IRX-\beta$ relation accounts for the scatter in the intrinsic $\beta$
from the effects of stellar population age (50~Myr to 1~Gyr), SFH (SFR$\sim
e^{-t/\tau}$, with 1~Gyr${~<~\tau~<~100}$~Gyr), and metallicity
(${0.02~Z_{\Sun}~<Z<~2.5~Z_{\Sun}}$).  The dashed lines show the
relations according to the parameterized dust law (see
\autoref{sec:DeltaMethod}) with (clockwise from left) $\delta$ = +0.4,
+0.2, 0.0, -0.2 -0.4.
} 
\label{fig:IRXcurves} 
\end{figure}

In \autoref{fig:IRXcurves}, several dust attenuation and extinction
laws from \autoref{fig:DustCurves} are shown on the plane of infrared
excess, $IRX$, and UV slope, $\beta$. Each dust law's $IRX-\beta$
relation represents the predicted location of a variety of stellar
populations that have been reddened according to their given
dust-attenuation or dust-extinction curve.  Creating these relations
requires several assumptions about the intrinsic stellar populations,
which manifest as an increase in the relation's width. First, we
obtained a library of BC03 stellar populations with a range of ages
(50~Myr to 1~Gyr), star-formation histories ({SFR$\sim e^{t/\tau}$}, where 
{$1~\text{Gyr}~<~\tau~<~100$~Gyr}), and metallicities (
{$0.02~Z_{\Sun}~<~Z~<~2.5~Z_{\Sun}$}). Then, we subtracted
the dust attenuated SED from the intrinsic SED and 
integrated the residual across all wavelengths to obtain an estimate of
the bolometric IR luminosity for
these models. We made the approximation that the calculated IR luminosity of each
model is representative of \lir, under the assumption that all attenuated
UV-to-NIR light is completely reprocessed to produce the total
IR luminosity.  
\referee{$L_{\mathrm{UV}}$ and $\beta$ for these models were calculated
using the same methods as used on the best-fit SEDs of the data (in
\autoref{sec:UV} and \autoref{sec:Beta}, respectively).} 


%
\setcounter{figure}{4}

From this framework, each dust model in \autoref{fig:IRXcurves} has a
width in the $IRX-\beta$ plane which is a product of the range in the
stellar population parameters (age, metallicity, star-formation
history), which affects both $IRX$ and $\beta$ and produce the scatter
illustrated by the colored swath.  The left edge represents younger,
low-metallicity, and maximally blue stellar populations, while the
right edge extends towards older, metal-rich, and intrinsically red
stellar populations.  With increasing steps of \ebv\ (moving up each
$IRX-\beta$ relation), a steeper dust law will redden the SED faster
and therefore produce less $IRX$ at a given $\beta$ when compared to
greyer, starburst-like dust laws \citep{Siana09}. In addition, the
presence of a 2175~\AA\ dust absorption feature, such as is found in
the Milky Way dust law, will produce a significant excess of IR
emission without significantly contributing to the reddening
\citep[although this depends on the manner in which $\beta$ is
determined, see][]{Kriek13}.  These $IRX-\beta$ relations provide an
observational basis with which to distinguish between dust-attenuation
curves.

\section{Distinguishing Between Dust Laws with Bayes Factors} \label{sec:BayesFactor}
Determining the shape of the dust-attenuation curve from broadband
data is nontrivial. Broadband SED fitting is fraught with parameter
degeneracies, a product of several physical mechanisms that conspire
to produce similar SED shapes \citep[\eg\ stellar population age,
metallicity, star-formation history, and dust attenuation; \eg][]{Papovich01,Papovich11,
JLee10,Lee11,Walcher11,Pacifici12,Pforr12,Pforr13,Mitchell13}.  As mentioned in
\autoref{sec:SEDfitting}, these degeneracies spawn biases in simple
$\chi^2$ likelihood-ratio tests because best-fit models are more
sensitive to SED template assumptions such as the inclusion of
nebular emission lines, changing the assumed dust curve and/or the
degeneracies within the parameters themselves \citep{Tilvi13,
Salmon15}. 

To distinguish between dust laws, we should
consider all parameters, $\Theta$, as nuisance parameters, such that
the fully marginalized parameter space contains probability
contribution from all $\Theta$. We are then left to quantify the
difference between the fully marginalized posteriors under their
respective assumptions of non-parametric dust laws. 
In order to achieve this, we consider the posterior in Bayes theorem
(\autoref{equ:Bayes}) as being further conditional to a model
assumption exterior to the fitting process (in this case, the assumed
dust-attenuation curve, $k_\lambda$).  Then we may determine the
odds that the hypothesis of one dust-attenuation curve is correct over
another. The ratio of a posterior assuming dust-attenuation curve
$k_\lambda^1$ and a posterior assuming dust-attenuation curve
$k_\lambda^2$ is therefore given by,
\begin{equation}
\label{equ:BayesRatio}
\begin{split}
\frac{P(\Theta',k_\lambda^1|D)}{P(\Theta',k_\lambda^2|D)} &= 
\frac{P(D|\Theta',k_\lambda^1)}{P(D|\Theta',k_\lambda^2)} \times
\frac{P(\Theta'|k_\lambda^1)}{P(\Theta'|k_\lambda^2)},  \\
\mathrm{or} \hspace{0.35 cm} \mathrm{posterior\ odds} &= \mathrm{Bayes\ factor}
\times \mathrm{prior\ odds}.
\end{split}
\vspace{0.0 cm}
\end{equation}

The term in the middle is referred to as the Bayes factor
\citep{Jeffreys35, Jeffreys61, Kass95}. In practice, we may write the
Bayes factor as a ratio of the marginal likelihood (see
\cite{Sutton01} for a similar definition). Combining the definition in
\autoref{equ:BayesianEvidence} with the conditions in
\autoref{equ:BayesRatio}, we obtain the plausibility that one
dust-attenuation curve is more likely given another, marginalized over all
parameters:
\begin{equation}
\label{equ:BayesFactor}
\mathrm{Bayes\ factor} \equiv\ B_{12}  =
        \frac{P(D|k_\lambda^1)}{P(D|k_\lambda^2)} 
\vspace{0.0 cm}
\end{equation}

\cite{Kass95} offered descriptive statements for Bayes
factors in order to denominate several standard tiers of scientific
evidence. These were defined using twice the natural logarithm of the
Bayes factor, which we will call the Bayes-factor evidence, $\zeta$:
\begin{equation}
\label{equ:BayesFactor2}
\begin{split}
\mathrm{Bayes} &\bdash{-factor\ evidence} \equiv\ \zeta = 2 \cdot \ln B_{12} \\
  \zeta   &= 2 \cdot \ln  
      \frac{\int_{\Theta} P(D|\Theta',k_\lambda^1)\ P(\Theta'|k_\lambda^1)\ \mathrm{d}\Theta}
           {\int_{\Theta} P(D|\Theta',k_\lambda^2)\ P(\Theta'|k_\lambda^2)\ \mathrm{d}\Theta} .
\end{split}
\vspace{0.0 cm}
\end{equation}
%

We adopt the significance criteria of \cite{Kass95}, who define the
evidence to be ``very strong'' ($\zeta > $10),
``strong" ($6 < \zeta <\ 10$), or ``positive" ($2\ < \zeta<\ 6$) 
towards $k_\lambda^1$ (and equivalent negative
values for evidence towards $k_\lambda^2$).  Intuitively, because
the Bayesian evidence $P(D)$ is proportional to the integral over the likelihood
(\autoref{equ:BayesianEvidence}), a model
that produces a better fit to the data (low $\chi^2$) will yield a
higher $P(D)$, making $|\zeta|$ larger in the case
that one dust law is more likely than another.

Throughout this paper, we refer to galaxies with high $|\zeta|$ as
having strong Bayes-factor evidence towards a given dust law.
However, we caution that Bayes factors do not necessarily mandate
which of two models is correct but instead describe the evidence
against the opposing model.  For example, a galaxy with very strong
evidence towards model~2 (\eg\ $\zeta \approx -20$ according to
equation~\autoref{equ:BayesFactor2}), promotes the \emph{null
hypothesis} that model~1 is correct. Formally, it does not say the
model~2 is the correct model, \referee{but promotes the rejection of
model~1} (and vice versa).  In the next section, we address this
subtlety with a direct parameterization of the dust-attenuation curve
in order to confirm if the Bayes-factor evidence is indeed pointing
towards the appopriate dust prescription. 

\section{Parameterizing the Dust Law} \label{sec:DeltaMethod}
While it is instructive to search for the evidence that galaxies have
one of the empirically or physically motivated dust laws from
\autoref{sec:DustCurves}, there is no guarantee that these dust laws
apply to all galaxies, particularly at high-redshifts.  We therefore
adopted an alternative model for the dust attenuation, where we
parameterize the dust law in the SED-fitting process. The
parameterization allows a smooth transition between
the different dust laws. Following \cite{Kriek13}, we allowed the
dust-attenuation curve to vary as a \referee{tilt} from the
starburst curve of \cite{Calzetti00} similar to the parameterization
provided by \cite{Noll09}. \referee{This paramaterized dust law,} 
which is a purely analytical interpretation of how the 
dust-attenuation curve may be adjusted is:
\begin{equation}
\label{equ:delta}
A_{\lambda, \delta} \equiv\
\ebv\ k_{\lambda}^{\mathrm{SB}}\ \left (\lambda/\lambda_{\text{V}}
\right )^\delta
\end{equation}
This definition returns the starburst attenuation curve,
$k_{\lambda}^{\mathrm{SB}}$, when $\delta=0$, a steeper, stronger
attenuation in the FUV when $\delta<$ 0, or a flatter, greyer
attenuation across UV-to-NIR wavelengths when $\delta>$ 0. Examples of
these dust laws are shown in \autoref{fig:DustCurves}. \referee{We
chose a parameter space with a range $-0.6<\delta<+0.4$ in steps of
$\Delta \delta$=0.1 (see Table~1). This range brackets the range of
dust laws observed in the literature.} In comparison, SMC92 is slightly steeper
than the starburst curve across UV-to-NIR wavelengths, similar to
$\delta \approx -0.1$, but is much steeper at $\lambda \lesssim
1500 \AA$, similar to $\delta \approx -0.5$. 

Given the set of dust laws, we can marginalize over all other
parameters $\Theta$ to obtain the posterior on $\delta$ for each
galaxy. This process is the same as the marginalization in
\autoref{equ:BayesianEvidence}, where we marginalize over all $\Theta$
to obtain the full marginal likelihood, except that we have added an
additional parameter $\delta$.  In some cases $\delta$ may be poorly
constrained, and the posterior will be very broad. This is to be
expected, as there is similarly a population of galaxies for which the
Bayes factor is unable to return significant evidence. The results of
fitting to $\delta$ are described in \autoref{sec:speczDeltaResults}
and \ref{sec:photzDeltaResults}.

\autoref{equ:delta} assumes there is no additional contribution from
the 2175~\AA\ absorption feature, which is a hallmark of the Milky Way
dust-attenuation curve \citep{Gordon03} and is likely caused by
absorption from polycyclic aromatic hydrocarbons.
Although there is evidence of the 2175~\AA\ feature in high-redshift quasars
\citep{Noterdaeme09}, gamma ray burst host galaxies
\citep{Elliasdottir09}, and star-forming galaxies
\citep{Noll07,Buat11}, its strength and prevalence in distant galaxy
populations remains uncertain \citep{Buat12}.  \referee{
As we discuss below (\autoref{sec:photzDeltaResults}), we tested for indications of the
2175\AA\ feature in the dust law and found no substantive evidence for it
based on our model fits to the broadband data.  We therefore did not
include the 2175\AA\ feature in our modeling. Introducing it would add
another parameter to the dust law \citep[see][]{Kriek13}.}

\begin{figure*}[!t]  
\centerline{\includegraphics[scale=0.5]{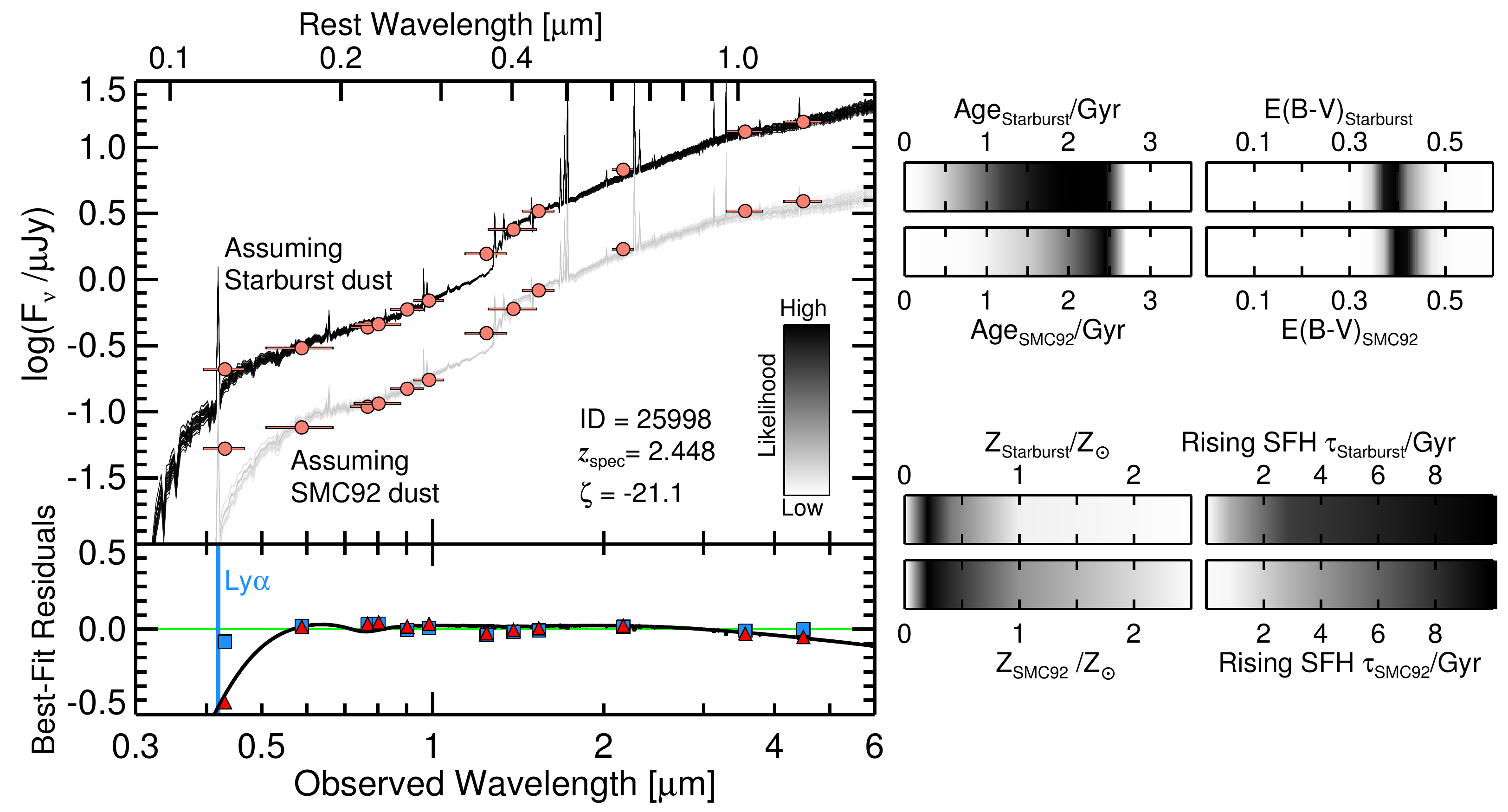}}
\caption{
The SED of a galaxy in our spec-$z$ sample with strong Bayes-factor
evidence ($\zeta=-21$) towards a starburst-like attenuation. The
salmon-colored photometric data points are shown twice, with the
second set offset by 0.6 dex for clarity.  The 50 most likely SEDs are
shown, scaled in opacity such that darker curves represent higher
likelihood up to the best-fit model, under an assumed starburst
(upper) or SMC92 (lower) dust law.  The black curve in the lower panel
shows the log-difference residual of the best-fit SED under each dust
assumption, and the average residual of the data and the 50 best-fit
starburst (SMC92) model fluxes in blue squares (red triangles).  The
residual of the best-fit SEDs are shown in the lower panel. The bars
to the right show the marginalized posteriors of individual
parameters, with darker regions denoting higher likelihood.  For
galaxies with very red SEDs across rest-frame $\lambda=0.2-2~\micron$
(the inferred \ebv\ is high) the UV-steep SMC92 dust law is incapable
of producing high-likelihood models that match both the
${B_{435}-V_{606}}$ color and the red rest-frame NIR color. 
This leads to the large difference in Bayesian evidence and the
low Bayes-factor evidence, $\zeta$.
} 
\label{fig:ExampleSEDCALZ} 
\end{figure*} 
\setcounter{figure}{5}
%


\section{The Non-Universality of Dust Laws at \zed\ $\sim$ 2} \label{sec:Results}

\subsection{Relevant Spectral Features} \label{sec:speczSEDresults}

\autoref{fig:ExampleSEDCALZ} shows the SED of a single galaxy in the
spec-$z$ sample that has strong Bayes-factor evidence promoting a
starburst dust-attenuation law. The SED features that drive the
differences in likelihood between the two dust assumptions are subtle.
In general, the rest-frame UV flux (1200~\AA\ $\lesssim \lambda
\lesssim$ 1400~\AA), which at the redshift range of this work is
either the $B_{435}$ or $V_{606}$ filter, catches the wavelength where the
dust laws differ the most. For galaxies like the example
in \autoref{fig:ExampleSEDCALZ}, the rest-frame optical-to-NIR SED
suggests a highly attenuated stellar population (high \ebv), yet
the flux from the rest-frame FUV band is brighter than the prediction
when assuming SMC92 dust. This results in a lower likelihood for SMC92 
models compared to models that assume starburst dust.
This is true even when accounting for the contribution from Ly$\alpha$
emission in the models or
variations to the assumed star-formation history. 
The likelihood difference, when marginalized over all parameters, is
reflected in the Bayes-factor evidence. 


\autoref{fig:ExampleSEDSMC} shows the SED of a single galaxy in the
spec-$z$ sample that has strong Bayes-factor evidence promoting an
SMC92 dust-extinction law. For this galaxy, the rest-frame
optical-to-NIR SED suggests a stellar population with relatively low
levels of attenuation (low \ebv). However, there is a subtle
decrease in the rest-frame FUV emission, which the starburst attenuation has
difficulty matching simultaneously with the rest of the SED, resulting
in less overall likelihood as compared to the SMC92 assumption.
Again, this likelihood difference is reflected in the Bayes-factor
evidence. 

One potential alternative explanation for the shape of these SEDs is a
two-component stellar population: a young burst of star formation
producing O- and B-type stars that dominate the rest-frame UV and an
already present intermediate-age population that dominates the
rest-frame optical-to-NIR SED.  As shown in the single-parameter
likelihood distributions of Figures~\ref{fig:ExampleSEDCALZ} and
\ref{fig:ExampleSEDSMC}, the exponential star-formation history is a
poorly constrained parameter with this dataset. Folding in additional
SFH parameters will require data with a higher wavelength resolution
of the SED in order to overcome its degeneracies with age,
metallicity, and the tilt and scale of dust attenuation.  

%
\begin{figure*}[!t] 
\centerline{\includegraphics[scale=0.5]{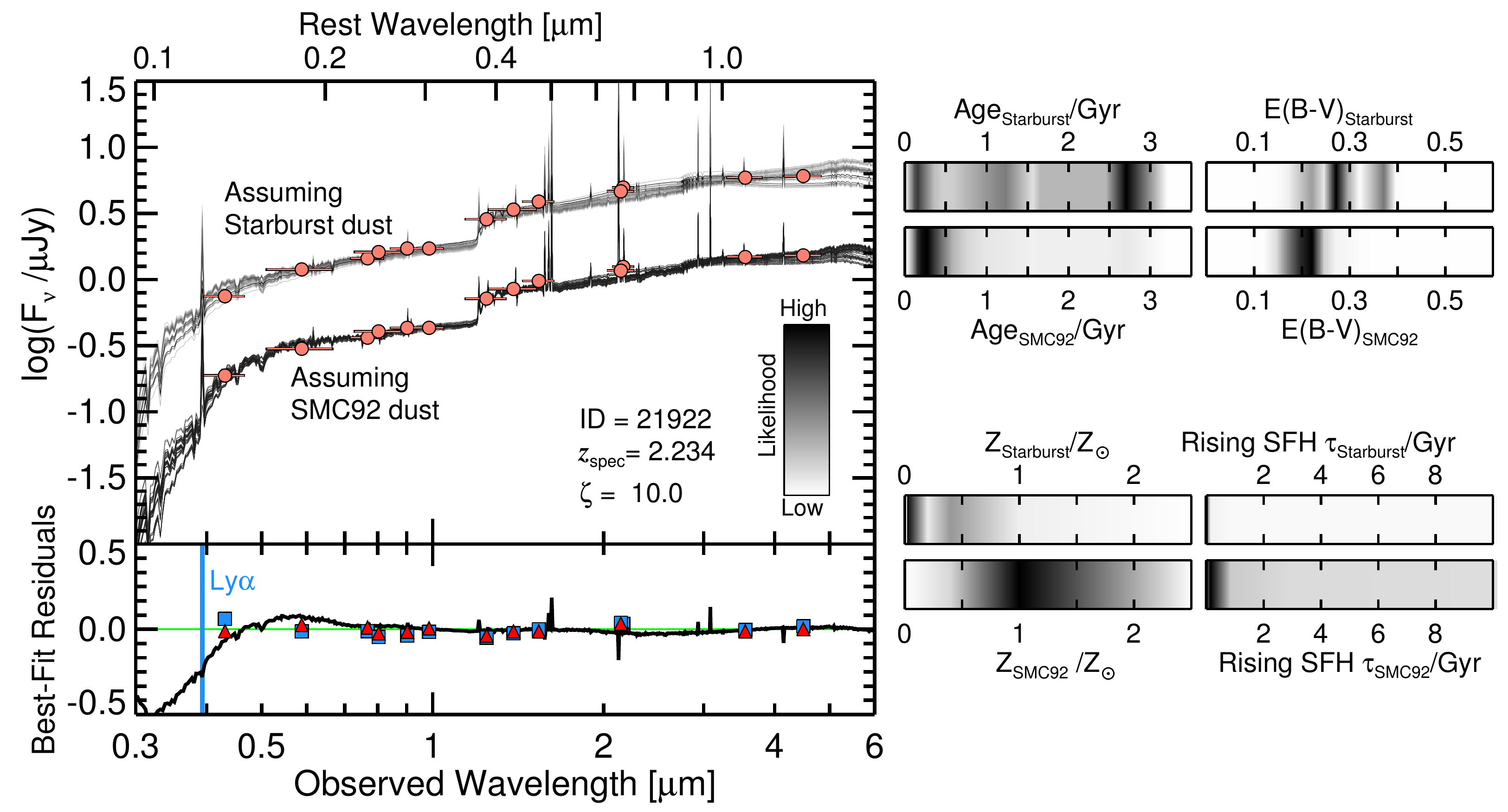}}
\caption{
\referee{The SED of a galaxy} that has strong
Bayes-factor evidence ($\zeta=10$) towards an SMC92-like
dust-attenuation law.  The data are shown twice for fits under each
dust assumption, offset by 0.6 dex for clarity. The results between
assuming starburst and SMC92 attenuations are subtle; when most of the
SED (rest-frame $\lambda =0.2-2~\micron$) suggests relatively low
values of \ebv, the assumpition of a starburst law does not produce as
many models with significant likelihood as the SMC92 law capable of
reproducing both the red ${B_{435}-V_{606}}$ color and the rest-frame
NIR color. The resulting difference in likelihood is reflected in the
Bayes-factor evidence, $\zeta$. 
} 
\label{fig:ExampleSEDSMC} 
\end{figure*} 
\setcounter{figure}{6}
%

\subsection{Results from the Spectroscopic Redshift Sample} \label{sec:speczResults}

\subsubsection{Bayes Factors on the $IRX-\protect\beta$ Relation:\\
spec-$z$ sample}\label{sec:speczIRX}

\autoref{fig:money} shows the selection of Bayes-factor evidence for
individual galaxies as a function of stellar mass.
As expected, most galaxies lack enough
evidence from their broadband data alone to distinguish their
underlying dust law. However, there are examples of galaxies that
display strong evidence towards having an SMC-like or starburst-like
attenuation.  
\autoref{fig:money} also shows the 
plane of $IRX-\beta$, where the total infrared
luminosities were calculated as described in \autoref{sec:lir} and
$\beta$ in \autoref{sec:Beta}. As noted in Figures
\ref{fig:ExampleSEDCALZ} and \ref{fig:ExampleSEDSMC}, the band closest
to the Ly$\alpha$ is the most sensitive to determining the evidence
towards a given dust law because it is at the wavelength where the dust
prescriptions differ the most.  

The Bayes-factor evidence for
different dust laws among is consistent with the galaxies' positions in the $IRX-\beta$ plane.
The Bayes-factor evidence was derived from the rest UV-to-NIR
photometry and shows that some galaxies have very strong evidence
against the starburst law or SMC92 law.   Those same galaxies have
$IRX-\beta$ measurements \referee{consistent} with the Bayes-factor evidence.
This is significant because the \lir\ data provide an independent
measure of the dust law.

Though the results of \autoref{fig:money} seemingly identify galaxies
with two types of underlying dust scenarios, we must recognize the
possibility that neither dust-attenuation curve is appropriate, even
for some of the \referee{objects with the strongest evidence.} In the next section, we
pursue this possibility using the methods described in
\autoref{sec:DeltaMethod} to parameterize the dust-attenuation curve
as a new variable in the fitting process. \\

\subsubsection{Fitting the Curve of the Dust Law:\\ spec-$z$ sample} \label{sec:speczDeltaResults}
\autoref{fig:BF_MedianDelta} shows the results of fitting to the
parameterized dust-attenuation curve (\autoref{equ:delta}).  The
selection of galaxies with strong evidence towards a starburst-like
dust-attenuation curve agrees with the results from fitting to the
dust-attenuation curve directly.  Similarly, SMC-like galaxies are
better described by a steeper dust-attenuation curve ($\delta<-0.2$),
albeit at varying degrees.  The galaxies selected to have strong
evidence towards an SMC92 dust law exhibit a marked steepness in their
fitted dust-attenuation curve that contrasts with a starburst
dust law. 

%
\begin{figure*}[!t] 
\centerline{\includegraphics[scale=0.64]{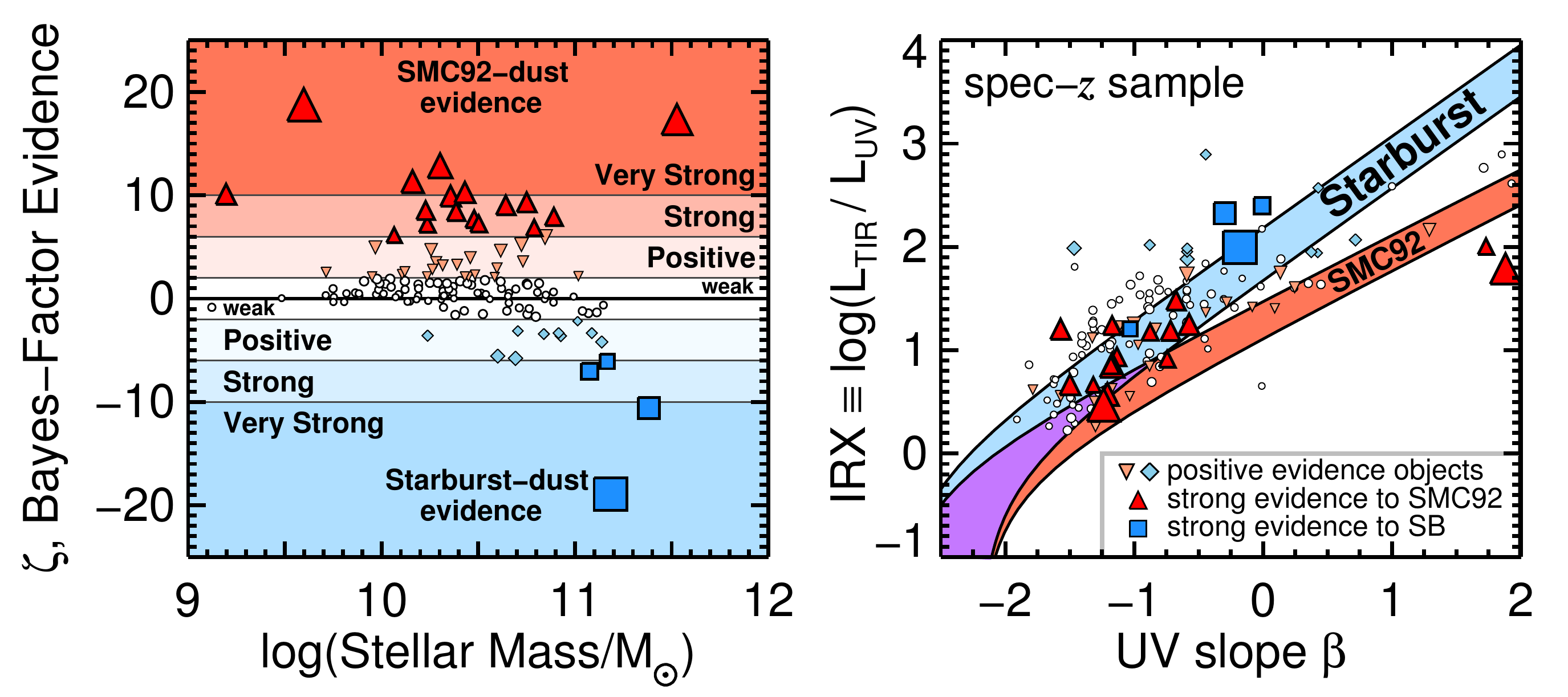}}
\caption{ 
Left: The Bayes-factor evidence as a function of stellar mass for
galaxies in the spec-$z$ sample with 1.5$\le$$z$$\le$2.5 and
MIPS~24\micron\ detections of S/N$>$3.  \referee{Darker} shaded regions
indicate levels of increasing evidence and are used to select objects
that show strong preference between SMC92 (red triangles) or starburst 
(blue squares) dust curves.  No mid-IR information was used in the
left figure; these values were achieved by modeling rest-frame
UV-to-NIR fluxes only.  Right: Measured IR excess versus UV slope 
to test results inferred from the UV/NIR SED.  Prediction curves
from stellar population models for SMC92 (red) and starburst (blue)
dust laws are shown. Objects selected by the strength of
their Bayes-factor evidence follow the curve of their predicted
dust-attenuation curve with some scatter.  The independent
measurements of the $IRX-\beta$ relation supports the Bayesian
evidence from the modeling of the UV/NIR SED:  galaxies with
(very-)strong Bayes-factor evidence follow the correct $IRX-\beta$
relation. 
} 
\label{fig:money} 
\setcounter{figure}{7}
\end{figure*}
\begin{figure}[!h]  
\centerline{\includegraphics[scale=0.45]{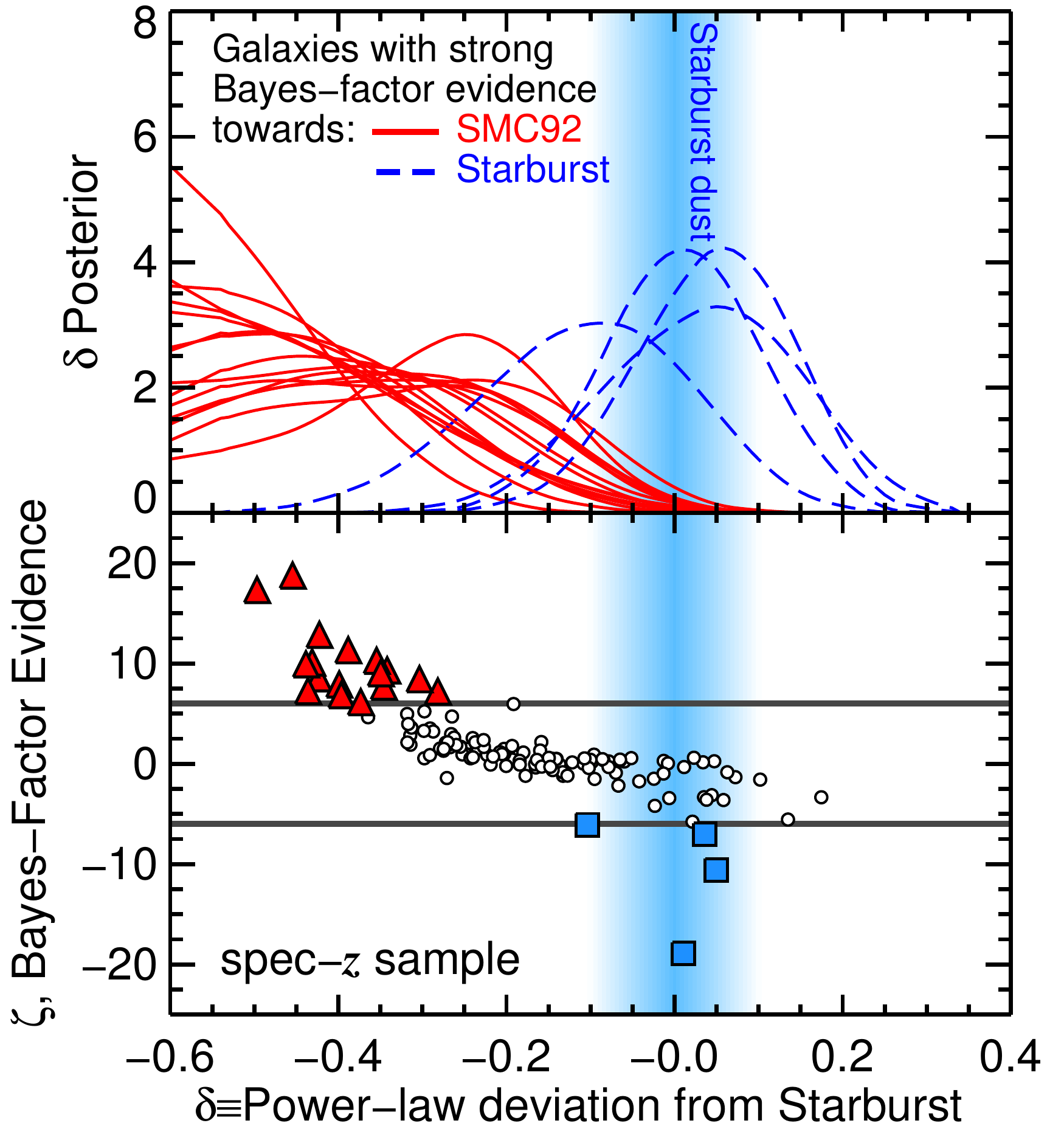}}
\caption{
Top: The posterior probability of the fitted parameter $\delta$, the
power-law deviation from a starburst \citep{Calzetti00} dust-attenuation curve,
for galaxies in the spec-$z$ sample and a broadband filter near
Ly$\alpha$. Each curve represents a galaxy that was selected in
\autoref{fig:money} as having strong Bayes-factor evidence towards an
SMC-like (red, solid) or a starburst-like (blue, dashed) dust-attenuation curve.
The width of the blue hazed region shows the typical 1-$\sigma$ uncertainty in
the median value of $\delta$, centered on $\delta=0$ where a galaxy
would have a starburst dust-attenuation curve.  Bottom: The evidence from the
Bayes factors between the SMC92 and starburst dust laws as a
function of the $\delta$ posterior median. Symbols shapes and colors
are the same as \autoref{fig:money}.  The Bayes factors of galaxies
with strong evidence broadly agree with the median $\delta$, as would
be expected. 
} 
\label{fig:BF_MedianDelta} 
\setcounter{figure}{8}
\end{figure} 

\begin{figure}   
\centerline{\includegraphics[scale=0.54]{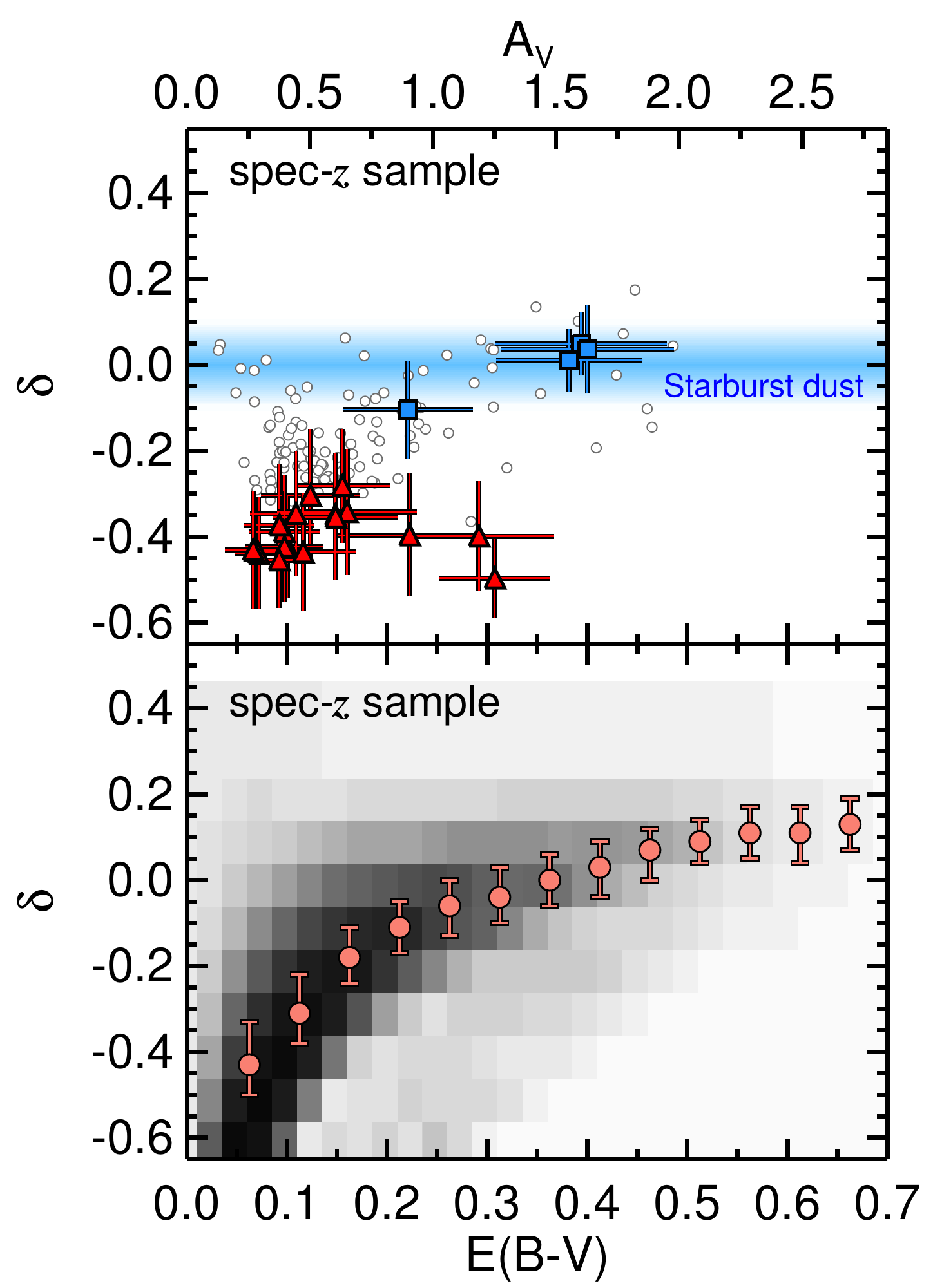}}
\caption{
Top: The median of the posterior on the \referee{tilt of the} dust-attenuation curve,
$\delta$, as a function of the median color excess, \ebv, for
galaxies in the spec-$z$ sample. Red triangles and blue squares are
galaxies with strong Bayes-factor evidence towards SMC-like and
starburst-like dust laws, respectively.  Galaxies with low
Bayes-factor evidence are shown as open white circles. For reference,
a blue haze is shown where $\delta$ represents a starburst
attenuation, where the width represents typical 1-$\sigma$
uncertainties on the posteriors of $\delta$.  Bottom: The joint
probability between the $\delta$ and \ebv\ posteriors.  Medians in
bins of \ebv\ and their 68\% limits are shown as salmon-colored points and error
bars respectively. The spec-$z$ sample suggests a relation between the
scale of dust attenuation and the \referee{tilt} of the dust law, such that 
galaxies with higher attenuation have a flatter (or grey)
starburst-like dust law, and galaxies with relatively lower
attenuation have a steeper, SMC-like dust law. 
} 
\label{fig:speczDelta} 
\setcounter{figure}{9}
\end{figure} 

In \autoref{sec:BayesFactor}, we mentioned how the Bayes factor is
formally promoting the \emph{null hypothesis} of the opposing model.
For example, the Bayesian evidence formally does not favor model~1,
but provides evidence against the competing model~2 compared to
model~1.   However, taken together, the results in
\autoref{fig:BF_MedianDelta} imply that galaxies with negative
$\delta$ really do have
steeper attenuation curves like that of the SMC92. In this case, we may
consider the evidence towards the null hypothesis of the opposing
dust law as being the same as positive evidence for the
hypothesis of the dust law itself.

One of the main results of this work is seen in
\autoref{fig:speczDelta}:  there is a strong relation between \ebv\
and $\delta$.  \autoref{fig:speczDelta} shows the derived values of
\ebv\ and $\delta$ for galaxies with high Bayes-factor evidence.
Because both axes are derived quanties with associated posteriors, we
combine the posteriors into a two-dimensional posterior for the whole
sample. In both cases, a clear trend emerges such that galaxies with
steeper, SMC-like dust laws also have lower levels of attenuation,
whereas galaxies with high attenuation have greyer, starburst-like
dust laws. This correlation agrees with the $IRX-\beta$
relation in \autoref{fig:IRXcurves}; galaxies with low $IRX$ are
expected to have steeper dust laws.  \\


%

%
%

\subsection{Results from the Photometric Redshift Sample}\label{sec:photzResults}

\subsubsection{Bayes Factors on the $IRX-\protect\beta$ Relation: \\
phot-$z$ sample}\label{sec:photzIRX} 
\autoref{fig:photzmoney} shows
the $IRX-\beta$ plot for the phot-$z$ sample (see
\autoref{sec:redshifts}). The phot-$z$ sample includes galaxies from
the spec-$z$ sample but with their redshifts assigned to their
photometric-redshift value.  \autoref{fig:photzmoney} also shows the
results from directly substituting the photometric redshifts for the
spec-$z$ sample, in order to explore how photometric-redshift accuracy
can affect the main results. The calculation of the UV
slope is also sensitive to the photometric-redshift uncertainty
because the bands used to find the slope may differ for large changes
in redshift (see the details on the calculation of $\beta$ in
\autoref{sec:speczIRX}).  It is plausible that galaxies in the
spec-$z$ sample have better photometric-redshift accuracies than those
for the full phot-$z$ sample.  However, we assume that the selection
bias to the right panel of \autoref{fig:photzmoney} is negligible
because the trends of Bayes-factor evidence on the $IRX-\beta$ plane
are the same for the phot-$z$ sample. 

\autoref{fig:photzmoney} shows that the phot-$z$ sample of this work is able to
reproduce the main result derived for the spec-$z$ sample.  In most cases, the
Bayes-factor evidence promotes the same dust law that the observations
suggest, based on their location in the $IRX-\beta$ plane.
However, there is significant scatter on an individual galaxy basis,
especially for the galaxies that seemingly promote an SMC92
attenuation (or discredit the starburst attenation). This is to be
expected; it is unlikely that all galaxies divide into two specific
types of dust laws. For example, the SMC92-favored galaxies may
have a range of attenuations that are, in different ways, steeper than
the starburst dust law.  \\

%
\begin{figure*}  
\centerline{\includegraphics[scale=0.64]{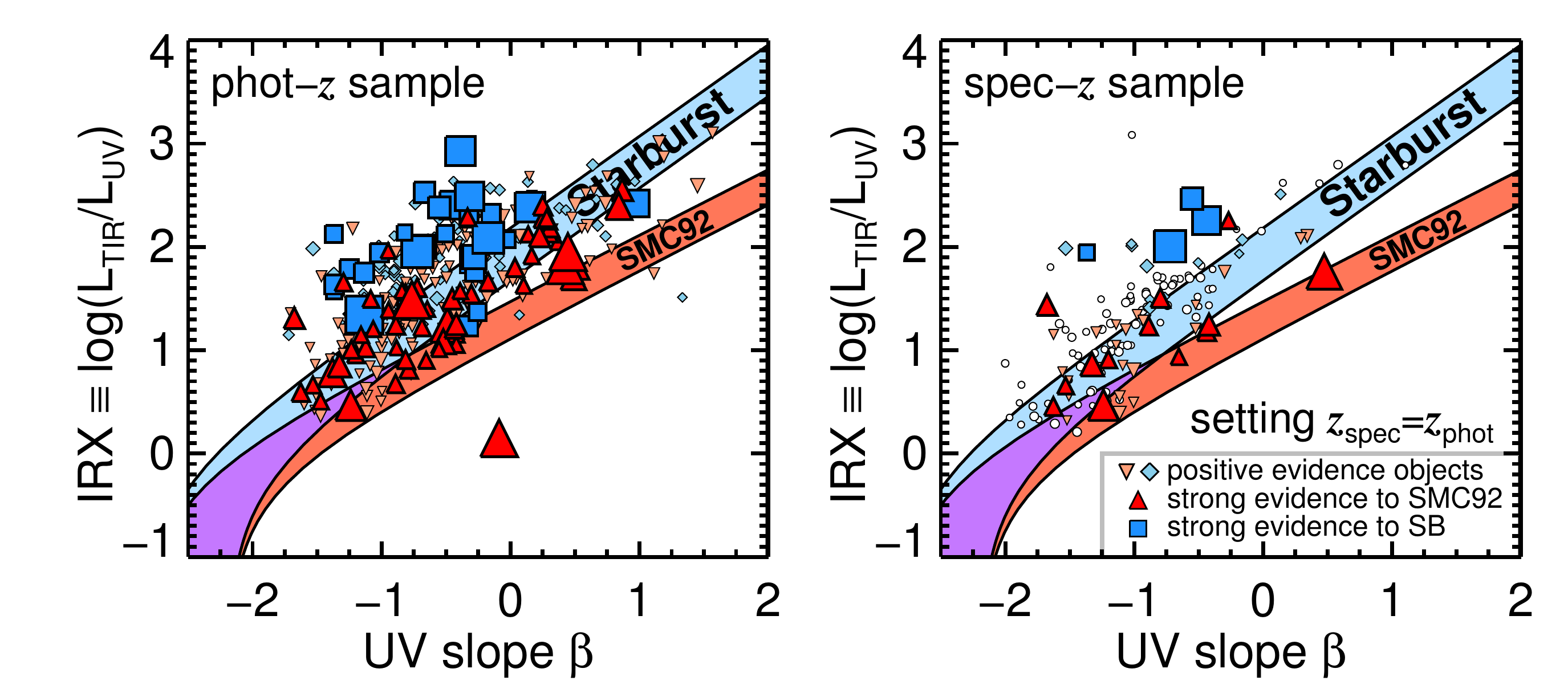}}
\caption{
Left: \referee{The measured IR excesss versus UV slope} for the
phot-$z$ sample (see \autoref{sec:redshifts}). For clarity, only galaxies
with positive and strong Bayes-factor evidence are shown, with the
symbol size scaling with the evidence. Curves show the predicted
location of a variety of stellar populations according to an SMC92 or
starburst dust-attenuation law. This panel shows that the galaxies in
the phot-$z$ sample selected by the strength of their Bayes-factor evidence
follow the curve of their predicted dust-attenuation law with some
scatter Right: The same as the right panel of \autoref{fig:money}
(galaxies from the spec-$z$ sample), but with $\beta$ and the
Bayes-factor evidence recalculated when the photometric redshift is used 
for the spec-$z$ sample. This panel shows that the selection methods and
Bayes-factor evidence can overcome the errors from photometric-redshift estimates
to predict galaxy dust laws, verified by their position
in the $IRX-\beta$ plane. 
} 
\label{fig:photzmoney}
\setcounter{figure}{10}
\end{figure*} 

\subsubsection{Fitting the Curve of the Dust Law: \\ phot-$z$ sample} \label{sec:photzDeltaResults}
\autoref{fig:photzDelta} shows the Bayes-factor evidence of the
galaxies in the phot-$z$ sample as a function of their $\delta$
posterior median. The median \referee{tilt} of the dust-attenuation
curve, marginalized over all combinations of stellar population age,
metallicity, and \ebv, agrees with the trends suggested by the
Bayes-factor evidence.  Galaxies with high SMC92 evidence tend to
allocate their likelihood around steeper \referee{tilts to the}
attenuation law ($\delta<-0.2$), and galaxies with high starburst 
evidence allocate towards shallower \referee{tilts to the} 
attenuation law ($\delta>0$). 

\begin{figure}  
\vspace{0.25cm}
\centerline{\includegraphics[scale=0.54]{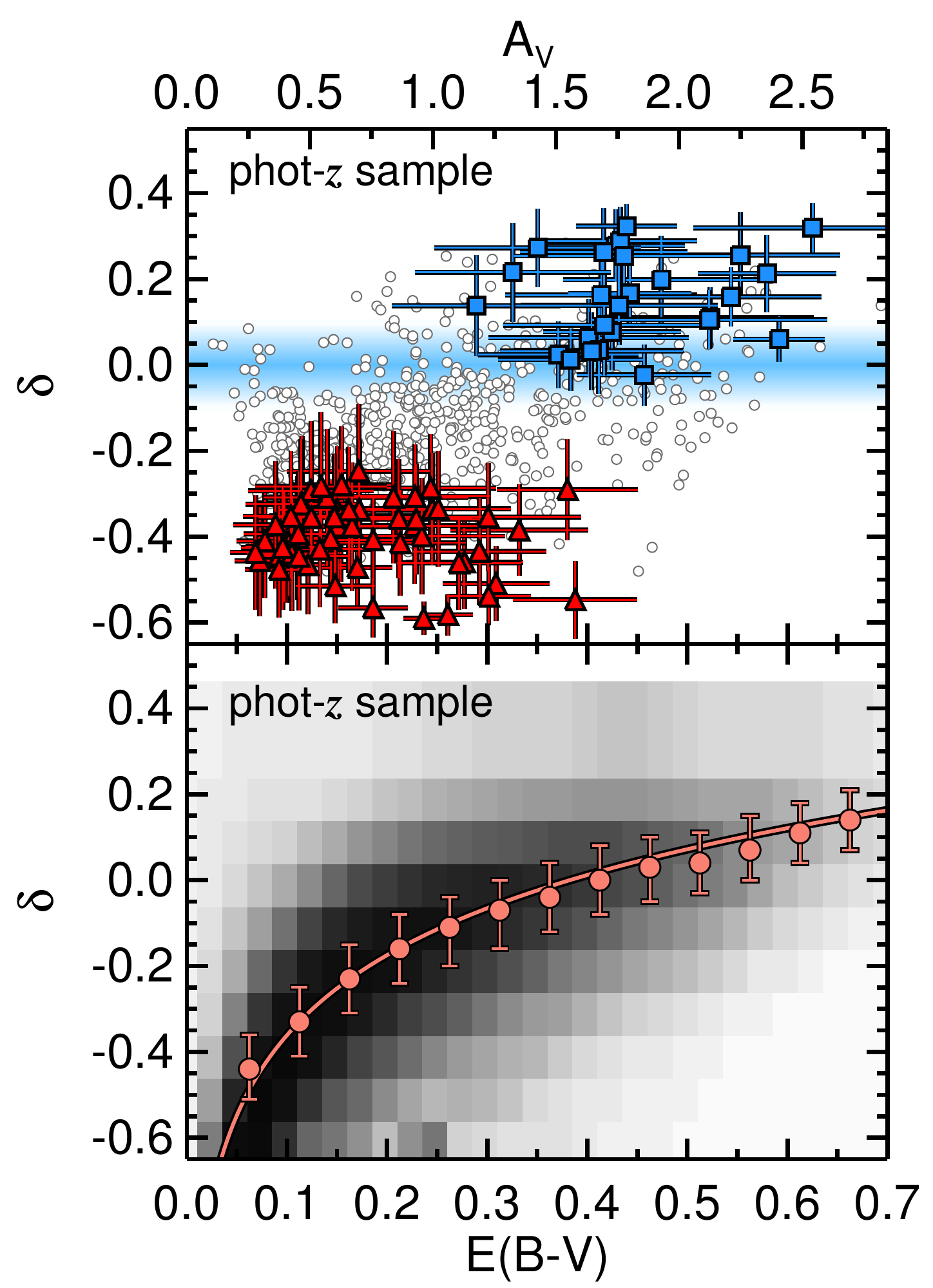}}
\caption{
The same as \autoref{fig:speczDelta} but for the phot-$z$ sample. Top:
The medians of the $\delta$ and \ebv\ posteriors are shown as grey
circles, with strong Bayes-factor-evidence galaxies highlighted as red
triangles (SMC92-like) and blue squares (starburst-like). For
reference, a blue haze is shown where $\delta$ represents a starburst
attenuation, given typical uncertainties in $\delta$.  Bottom: The
joint probability between the $\delta$ and \ebv\ posteriors.  The
salmon-colored circles show the $\delta$ at median likelihood in bins
of \ebv\ and error bars represent their 68~\% range in     likelihood.
The solid line represents a fit to the medians following
\autoref{equ:EBVdelta}.
} 
\label{fig:photzDelta}
\setcounter{figure}{11}
\end{figure} 
\autoref{fig:photzDelta} again shows that the
results of the phot-$z$ sample are an extension of the results from
the spec-$z$ sample (\autoref{fig:speczDelta}). This figure shows
that the steepness of the dust-attenuation curve correlates with the 
galaxy's attenuation optical depth, as parameterized by the color
excess. Galaxies that seem to scatter away from the main trend 
have poor wavelength coverage of rest-UV wavelengths and
have relatively broad posteriors in \ebv\ and $\delta$. 

\autoref{fig:photzDelta} also shows the posterior joint probability
between $\delta$ and \ebv\ for all galaxies in the phot-$z$ sample.
In this depiction, the galaxies with poor constraints on $\delta$ or
\ebv, which appear as outliers according to their median posteriors, 
get suppressed relative to the trend of the whole sample. 
The distribution shows a probability covariance such
that galaxies with low attenuation optical depths have steeper
dust laws and are well-fit by the relation 
\begin{equation} \label{equ:EBVdelta}
{\delta = ( 0.62 \pm  0.05) \log(E(B-V)) + 0.26 \pm  0.02}
\end{equation}

\referee{
We tested for the effect of the 2175\AA\ dust feature by refitting all
the galaxies but excluding any band with a central
rest-frame wavelength within $\pm$250~\AA\ of this feature. This
excludes at most one band for each galaxy. The relation in \autoref{fig:photzDelta}
was unchanged, implying that the dust feature does not affect our
results.}
\section{Discussion} \label{sec:Discussion}
\subsection{Origins of the relation between $E(B-V)$ and $\delta$}
\referee{The parameter $\delta$ applies a spectral tilt to the
attenuation law that pivots about the central wavelength of the $V$
band.  Unlike the original definition by \cite{Noll09}, our
parameterization allows the
total-to-selective attenuation at the $V$ band, \Rv, to change.  This
is because \Rv\ is inversely proportional to the slope of the dust law
around ${\lambda=\lambda_\text{V}}$, which means that low $\delta$
implies low \Rv. \referee{This is physically motivated by the fact
that different dust laws have different $R_V$. For example, $R_V=2.95$
for the SMC dust law \citep{Pei92} and $R_V=4.05$ for the starburst
dust law \citep{Calzetti00}.}
Therefore, while the relation of \autoref{fig:photzDelta} displays a
correlation between $\delta$ and \ebv, it could be interpreted as a correlation
between \Rv\ and \ebv. Intriguingly, this result may be the
consequence of dust physics in galaxies. The value of \Rv\ has been
linked to the average dust grain size
\citep{Gordon00}, with smaller grains being associated with smaller
\Rv. However, even if the result in \autoref{fig:photzDelta} was
linked to a change in dust grain size, it would still be difficult to
comment on the dust production sources, given the short timescale for
dust grain evolution \citep{JonesAP13}. 
}


\referee{While there is some covariance between $\delta$ and \ebv\ in
the posteriors from the model fits, this does not drive the observed
correlation between them.  The covariance between the parameters can
be understood as follows.  Imagine} an SED well represented by some
$\delta$ and then applying a very small increase in $\delta$ towards a
flatter dust curve, keeping other parameters fixed. This would produce
less attenuation to UV bands, and the models respond by applying more
likelihood to higher values of \ebv\ in order to attenuate the UV.
While this is a known parameter degeneracy that is unavoidable with
the current data set, \referee{we explore a variety of tests in
\autoref{sec:AppendixDelta} to confirm that the relation in
\autoref{fig:photzDelta} is real despite the influence of covariance.
In addition, \autoref{fig:money} serves as independent, observational
confirmation} that the \referee{tilt} of the dust law correlates with
the amount of attenuation. 

%
\begin{figure} %
\centerline{\includegraphics[width=0.49\textwidth]{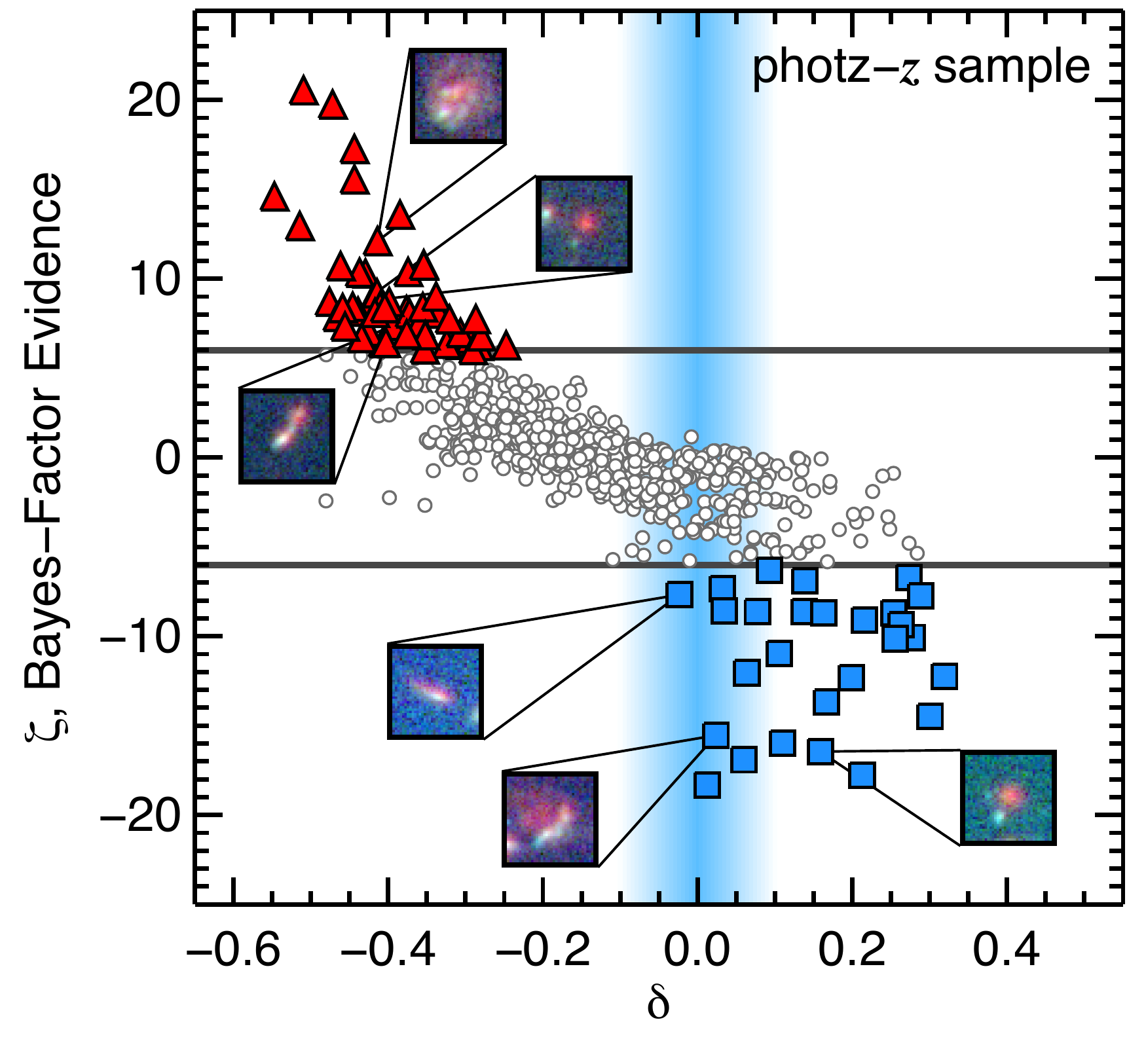}}
\caption{
Top: Bayes-factor evidence versus $\delta$ for the phot-$z$ sample.
Galaxies with strong Bayes-factor evidence are highlighted as red
triangles (SMC92-like) and blue squares (starburst-like). Example RGB
image stamps ($H_{160}$, $V_{606}$, and $B_{435}$, respectively) of
galaxies from both subsamples are shown as insets. This figure
demonstrates that there are no distinct morphological trends with
$\delta$ or Bayes-factor evidence. 
} 
\label{fig:stamps}
\setcounter{figure}{12}
\end{figure} 

\subsection{Physical Origins of Non-Universal Attenuation}
The ways that dust grain type, size, and distribution affect the
observed UV-to-NIR attenuation are enigmatic. There are many
physically motivated explanations for how the relationship between
stellar emission and the scattering and absorption by dust manifest to
an observed attenuation \citep{Witt00}. 
One possible explanation for the change in observed attenuation for
different galaxies is their orientation. There is evidence that galaxy
inclination correlates with the strength of Ly$\alpha$ emission, such
that we observe less Ly$\alpha$ equivalent width for more edge-on
galaxies \citep{Charlot93,Laursen07,Yajima12,Verhamme12,U15}. Resonant
scattering and absorption by dust are likely the primary impediments
to the escape of UV light from star-forming regions, which were
predicted by \cite{Charlot93} to be exacerbated in edge-on galaxies.
This qualitatively agrees with radiative transfer simulations that
show an increasing attenuation optical depth with galaxy inclination
\citep{Chevallard13}. Therefore, based on physical models, one expects
that galaxies with ``greyer" dust laws and larger overall attenuation
should have higher inclinations, on average.

However, we find no correlation with the scale or shape of attenuation
and the axis ratios in either the phot-$z$ or spec-$z$ samples.
\autoref{fig:stamps} shows the selection of galaxies in the phot-$z$
sample with strong Bayes-factor evidence. The inset image stamps are a
few examples that show similar morphologies among SMC92-like and
starburst-like galaxies. Compact red, large axis ratio, and clumpy
extended galaxies are found in both samples.  A more detailed study
with a wider mass range may be needed to find correlations with
inclination, axis ratio, or s\'{e}rsic index.  Alternatively, it may
be that neither \hst\ or \spitzer\ provides the wavelength coverage
with high enough angular resolution to discern the trends between
attenuation and morphology.  Future observations with \emph{JWST}
(with an angular resolution seven times higher than \spitzer\ at
similar wavelengths) may be needed to offer spatial insight on the
morphology of warm dust regions. 

Even if galaxy orientation/inclination correlates with the strength of
attenuation, it may not be the fundamental cause of non-universal
shapes to the dust-attenuation law. For example, \cite{Chevallard13}
predicted the relation between $\delta$ and attenuation optical depth
at all orientations, assuming only Milky-Way type dust grains. Their
model predicted a relationship between the shape of the extinction law
(here parameterized by $\delta$) and the dust-attenuation optical
depth.  We consider this relationship in two scenarios: small and
large dust-attenuation optical depths, $\tau$, where the optical depth
is related to the color excess by {$\tau_\lambda = 0.92\ k_\lambda\
E(B-V)$}.

In the low-attenuation scenario, the steep curve of the dust law is
likely a product of dust scattering, specifically the asymmetry
parameter of the scattering phase function and its dependence with
wavelength.  The asymmetry parameter, $g_\lambda$, describes the
degree of scattering in the forward direction \cite{Mann09}. Dust is
more forward scattering at UV wavelengths, such that $g_\lambda$
approaches unity, and more isotropic at optical-to-IR wavelengths,
such that $g_\lambda$ approaches zero
\citep{Gordon94,Witt00,Draine03b}. This means that in the
small-optical-depth regime, red light will tend to scatter
isotropically and escape the galaxy, while blue light will tend to
forward scatter until absorption. Therefore, relatively more
optical-to-IR light and less UV light escapes the galaxy, resulting in
a steepening of the curve of dust-attenuation ($\delta<0$). This only
applies at small optical depths where light has a chance to scatter
out of the galaxy before absorption.
\begin{figure}  \label{fig:Gordon}
\centerline{\includegraphics[scale=0.54]{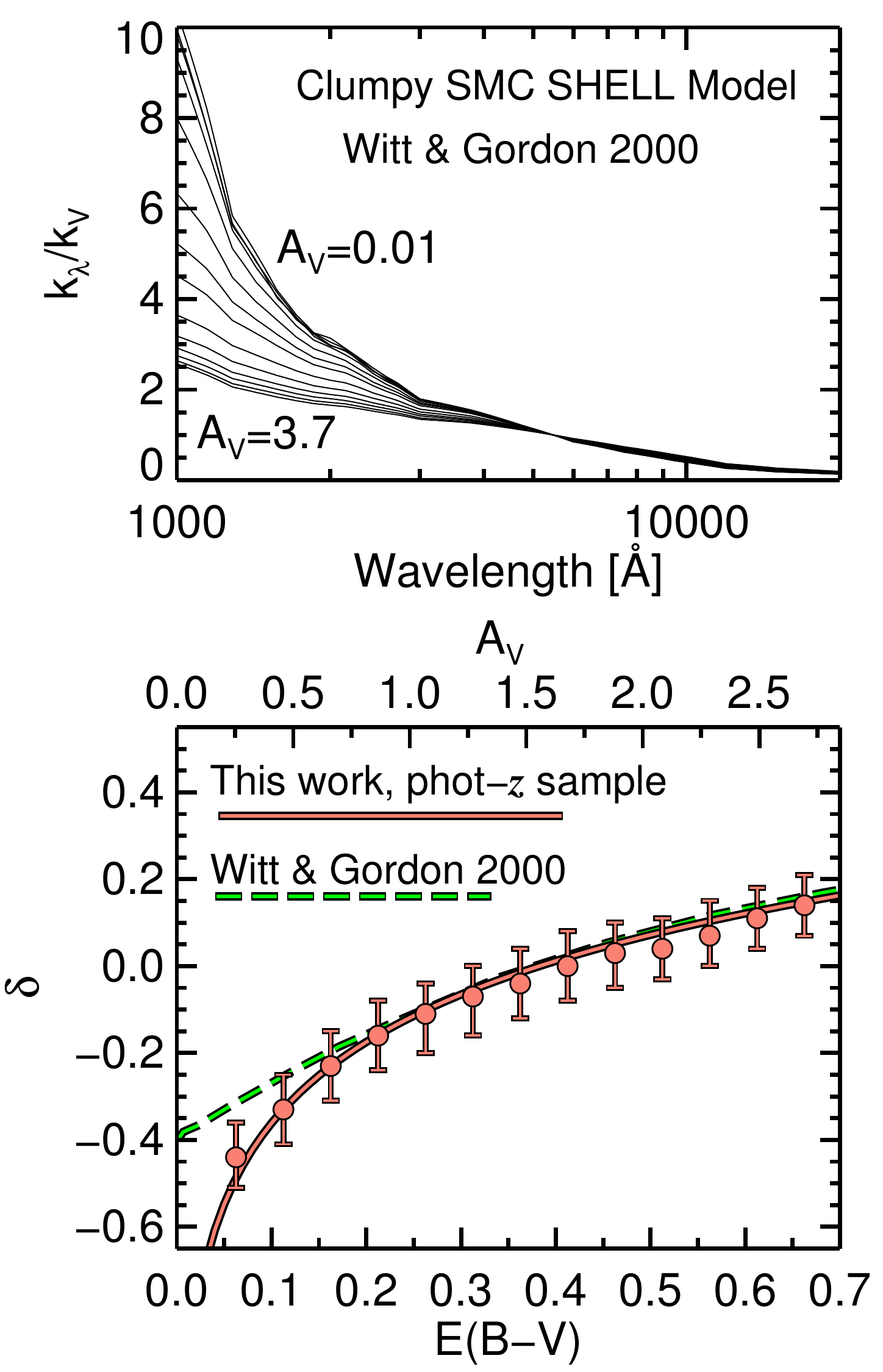}}
\caption{
Top: attenuation curves from the radiative transfer calculations by
\cite{Witt00} assuming SMC-like dust grains, a clumpy density
distribution, and a spherical shell geometry.  Bottom: The relation
between \ebv\ and $\delta$ from \autoref{fig:photzDelta}. The
salmon-colored circles show the $\delta$ at median likelihood in bins
of \ebv\ and error bars represent their 68~\% range in likelihood. The
solid line represents a fit to the medians following
\autoref{equ:EBVdelta}. The dashed green line represents radiative
transfer predictions from the above dust-attenuation models. 
}
\setcounter{figure}{13}
\end{figure}
Also in the low-attenuation scenario, dust is more transparent and
scattering is less frequent, so the scattering asymmetry parameter may
not be the only source of a steeper dust law. Galaxies with smaller
dust optical depths may have steeper dust laws because they produce
less scattering into the line of sight, causing the galaxy's
dust-attenuation law to appear more like a dust-extinction law.
\referee{In that case, the effects of dust grain size become more
pronounced.} For example, the steepness of the SMC extinction law has
been attributed to its observed underabundance of carbon, which
implies fewer heavy-element graphite grains than \referee{smaller}
silicate interstellar grains \citep{Prevot84}. This picture is
consistent with the trend of finding galaxies with more SMC-like dust
at very high redshifts \citep[\eg\ $z>5$ by][]{Capak15}, where the
metallicity of galaxies, in a broad sense, is expected to be lower
\citep{Madau14}.  The low $\delta$ for small \ebv\ in this work can,
at least in part, be attributed to lower \Rv. Thus, the relation
between \ebv\ and $\delta$ could be a product of underlying relations
in \referee{grain size,} averaged over the surface brightness of the
galaxy.  

In the regime of large attenuation optical depths, attenuation becomes
ubiquitous with wavelength. The flatter curve of the dust-attenuation
law resulting from high attenuation is consistent with the picture
proposed by \cite{Charlot00} whereby galaxies have a mixed
distribution of stars and dust. In this case, any escaping UV light
must come from regions of small optical depth, which corresponds to UV
light at the outer ``skin" of the mixed distribution
\citep{Calzetti01}. Conversely, redder light will come from deeper
physical locations within the region. The resulting attenuation
function is grey, or flatter with wavelength, which translates to
$\delta \geq 0$. 

\subsection{Comparison with Dust Theory}
\autoref{fig:Gordon} shows the predictions from radiative transfer
calculations by \cite{Witt00}. In general, the curves of
dust-attenuation become greyer at increasing optical depths for models
assuming SMC-like dust grains, a clumpy density distribution, and a
spherical shell geometry.  We determined $\delta$ for each curve using
the definition in \autoref{equ:delta} and compared its evolution with
\ebv\ to the results of \autoref{fig:photzDelta}. The steepest curves
in \autoref{fig:Gordon} are poorly represented by the $\delta$
parameterization (their curvature is higher than a power-law can
reproduce), which explains the disagreement between \ebv\ and $\delta$
at the low end.  The radiative transfer relations at high and low
optical depths are consistent with the observed correlation between
\ebv\ and $\delta$ found in this work. 

Although the agreement in \autoref{fig:Gordon} seems obvious given the
prevalent predictions of dust theory \citep{Bruzual88, Witt92,
Gordon01, Charlot00}, radiative transfer simulations \citep{Witt00,
Gordon00, Chevallard13}, and observations of local nebular regions
\citep{Draine01,Draine03b}, this is the first time the trend has been
found from only UV-to-NIR broadband photometry of distant galaxies.
Indeed, investigating origins of the relation between \ebv\ and
$\delta$ elucidates provocative explanations, as a result of similar
correlations in stellar population age, metallicity, and dust grain
\referee{size} from dust theory.

\subsection{Comparisons with Recent Literature} 
Several studies have noted populations of galaxies that lie off of the
nominal \cite{Meurer99} $IRX-\beta$ relation, suggesting galaxies with
younger ages have steeper, SMC-like dust laws
\citep{Siana09,Reddy06,Reddy10,Reddy12a,Buat12,Sklias14}.  Other
recent studies find galaxies at high redshift harbor dust laws that,
at least in large subsets of their samples, agree with the assumption
of a starburst dust law \citep{Scoville15, deBarros15, Zeimann15,
Kriek13}. Our results suggest that these studies are not in conflict
but provide clues to the overall non-universality of how dust
attenuates light in star-forming galaxies.  For example, the studies
that find evidence for starburst-like dust laws do so with galaxies
selected by their strong nebular emission lines (\eg\ H$\alpha$ and
H$\beta$), by recent star-formation activity (strong CIV absorption),
and/or by SEDs with appreciable reddening, allowing the underlying
dust law to be tested.  A common thread to these starburst-dust
galaxies is their large dust-attenuation optical depths, assuming
galaxies with strong nebular features also have high levels of stellar
attenuation by dust, averaged over the whole galaxy as foreground
screen \citep[see][for a discussion on alternatives to the foreground
screen]{Penner15}. 

The trend with $\delta$ and $E(B-V)$ (or with $IRX$) was also found by
\cite{Buat12} and \cite{Kriek13}. The latter authors attributed the
change in $\delta$ to a change in the strength of H$\alpha$ equivalent
width. The difference in this study is that the evolution of $\delta$
is determined for individual galaxies based on rest-frame UV-to-NIR
broadband photometry, suggesting that the methods can be used at even
higher redshifts. 

\begin{figure}  \label{fig:UVJ}
\centerline{\includegraphics[scale=0.57]{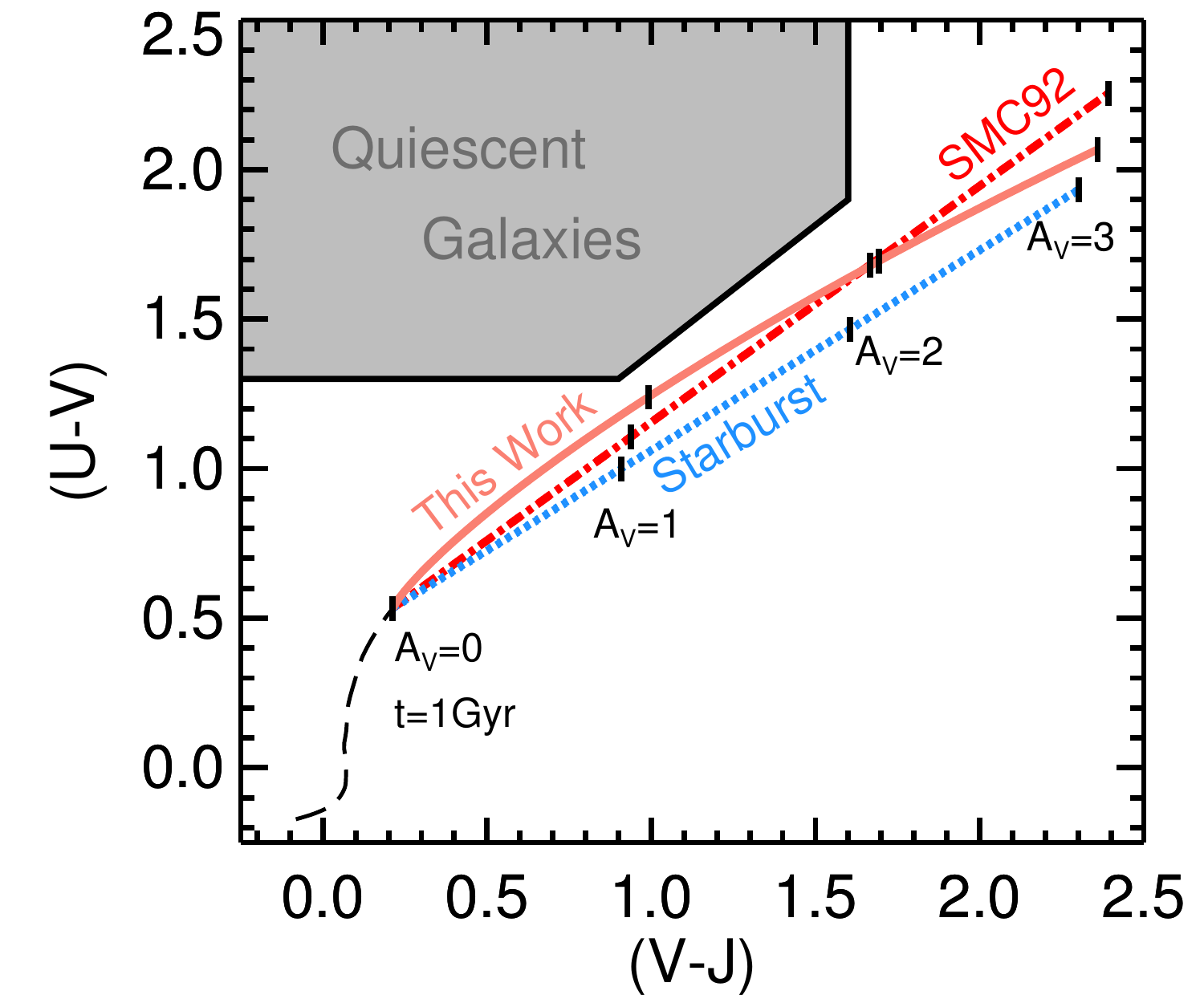}}
\caption{
Tracks of dust attenuation under different dust laws on the $UVJ$
diagram. The dashed black line follows the age of a stellar population
with a constant star-formation history from $t=0.02$ Gyr to $t=1$ Gyr.
The colored lines show how the rest-frame $UVJ$ colors of the $t=1$
Gyr stellar population change with increasing levels of attenuation
($0<A_\text{V}<$3) following a starburst-like dust law (blue, dotted), an 
SMC92-like dust law (red, dot-dashed), and a dust law that follows
\autoref{equ:EBVdelta} (salmon, solid). 
}
\vspace{0.2 cm}
\setcounter{figure}{14}
\end{figure}
\subsection{Implications for SED-derived properties of galaxies} 
The determination that individual galaxies at $z\sim2$ have different
dust laws has several implications for determinations of distant
galaxy evolution.  At relatively low \ebv\ ($\approx0.1$), this
manifests as a factor of $\approx$2 underprediction in the 1500~\AA\
luminosity dust correction (${10^{0.4 k_\lambda E(B-V)}}$) and
therefore UV SFR compared to the $\delta=0$ starburst assumption.
Higher SFRs for SMC92-like galaxies agrees with the determination of
stellar population ages: galaxies with an SMC92-like attenuation, are
on average half the age of galaxies with a starburst-like attenuation,
consistent with the results of previous studies \citep{Siana09,
Reddy10}. 
At \ebv$>$0.6, UV luminosity dust corrections are overestimated by a
factor of $2-5$ compared to the starburst-dust assumption at fixed
\ebv. 

\referee{
The dust law also has implications for studies that use the rest-frame
colors to infer dust attenuation in star-forming galaxies. The rest-frame $UVJ$
diagram is a particularly helpful visualization used to break the
degeneracy between old stellar population ages and reddening due to
dust \citep{Wuyts07,Williams09}.  Star-forming galaxies move along
the $UVJ$ diagram in an identifiable relation, where the redder colors
are to increasing levels of attenuation by dust \cite[][Fang et al.
2016 (in prep.)]{Price14,Forrest16} \autoref{fig:UVJ} shows how the
rest-frame $UVJ$ colors change with attenuation under different
assumed dust laws.  
The dust law derived in this work may induce slight differences to the
positions of star-forming galaxies, but these will be degenerate with 
star-formation histories and age/metallicity variations.
In addition, the dust law would not significantly influence the 
selection between star-forming and quiescent galaxies. 
}

Lastly, the relationship between $\delta$ and \ebv\ does not translate
to a relationship between the UV spectral slope $\beta$ and the dust
law.  At fixed $\beta$  there is high scatter in $IRX$ (see Figures
\ref{fig:money} and \ref{fig:photzmoney}).  Even at relative modest UV
slopes, {$-1< \beta < 0$}, the scatter in \lir/$L_{\text{UV}}$ is more
than 1 dex depending on the assumption of the dust attenuation law
\citep{Buat12}. This may have a substantive impact on the
interpretation of the intrinsic UV luminosity function if the
dust-corrections to the observed UV lumonsity densities assume a
unique relationship between $M_{\text{UV}}$ and $\beta$. \clearpage

\section{Conclusions} \label{sec:Conclusions}
We investigate the shape of the dust-attenuation law in star-forming
galaxies at $z\sim2$ in the CANDELS GOODS-N and GOODS-S deep fields.
We apply a Bayesian SED-fitting technique to galaxies with
spectroscopic and photometric redshifts, and determine the evidence
for their underlying dust law and its correlation with other galaxy
physical properties.  Our results can be summarized as follows:

\begin{itemize}
\item IR luminous galaxies at $z\sim$ 2 can be characterized by a
range of dust laws bounded by two types: (1) A starburst-like
\citep{Calzetti00} attenuation that is greyer (flatter) across
UV-to-NIR wavelengths and (2) a dust law that steepens towards the
rest-frame FUV like the curve of the SMC extinction law \citep{Pei92}. 
\item The dust law inferred from rest-frame UV-to-NIR photometry of
galaxies is supported by their position along the $IRX-\beta$
relations. This result gives credibility that a Bayesian analysis of
rest-frame UV-to-NIR fluxes is capable of broadly distinguishing
between dust laws that are grey or steep in the rest-frame FUV.  
\item The steepness of the dust law, parameterized by a $\delta$
power-law deviation from the starburst dust law, is correlated with
their color excess, \ebv, for IR-bright galaxies at $z\sim2$. Galaxies
with lower levels of dust attenuation have dust laws that are steeper
in the rest-frame FUV, following ${\delta = ( 0.62 \pm  0.05) \log(E(B-V)) + 0.26 \pm
0.02}$
\item The relation between \ebv\ and $\delta$ is further supported by
predictions from radiative transfer. The agreement with dust theory
offers possible interpretations for the relation and its origins from
different dust production mechanisms. For example, the change in the
shape of the dust law may be a result of star-dust geometry,
properties of dust grains, and/or stellar population age, emphasizing
the non-universality of the dust law in star-forming galaxies.
\end{itemize}

\vspace{-0.4 cm}
\appendix
\section{\referee{A. Testing the Relationship Between $E(B-V)$ and $\delta$}}\label{sec:AppendixDelta}
\referee{The main result of this paper is the relationship between the
tilt of the dust law $\delta$ and the amount of attenuation as probed
by the color excess, \ebv.  While this relationship is qualtitatively
and independently supported by the position of galaxies on the
$IRX-\beta$ relation and their Bayes-factor evidence, it is prudent to
consider if the covariance between $\delta$ and \ebv\ contributes to the
observed correlation (\autoref{fig:photzDelta}).
We addressed this concern with several tests below. }

\referee{First, we defined a grid of $\delta$ and $\ebv$ that span the
plane in \autoref{fig:photzDelta}. Then, we calculated fluxes from our
model SEDs for each value of $\delta$ and \ebv.  The model flux at
each bandpass was perturbed according to a Gaussian error distribution
where sigma was defined as the average, flux-dependent signal-to-noise
of the CANDELS data.  We used those fluxes as inputs to our procedure
and compared the derived values to the ``true" values of $\delta$ and
\ebv. We repeated this test fifteen times for five assumptions of the
input SED stellar population age and three assumptions of metallicity.
}

\referee{We find that when the input stellar population age is less
than a gigayear, we accurately recover all ``true" $\delta$ and \ebv\
across with no systematics or appreciable covariance and for all input
metallicities.  The covariance begins to appear when the galaxies are
older than 1 Gyr, and the effect increases in strength at higher
metallicities.  In that case, the posterior distributions for \ebv\
and $\delta$ are systematically shifted by $\sim$20\% of their
respective input values, where the covariance is positive and moves
models with higher $\delta$ to higher \ebv\ and vice versa.}

\begin{figure*}  \label{fig:AppendixDelta}
\centerline{\includegraphics[scale=0.52]{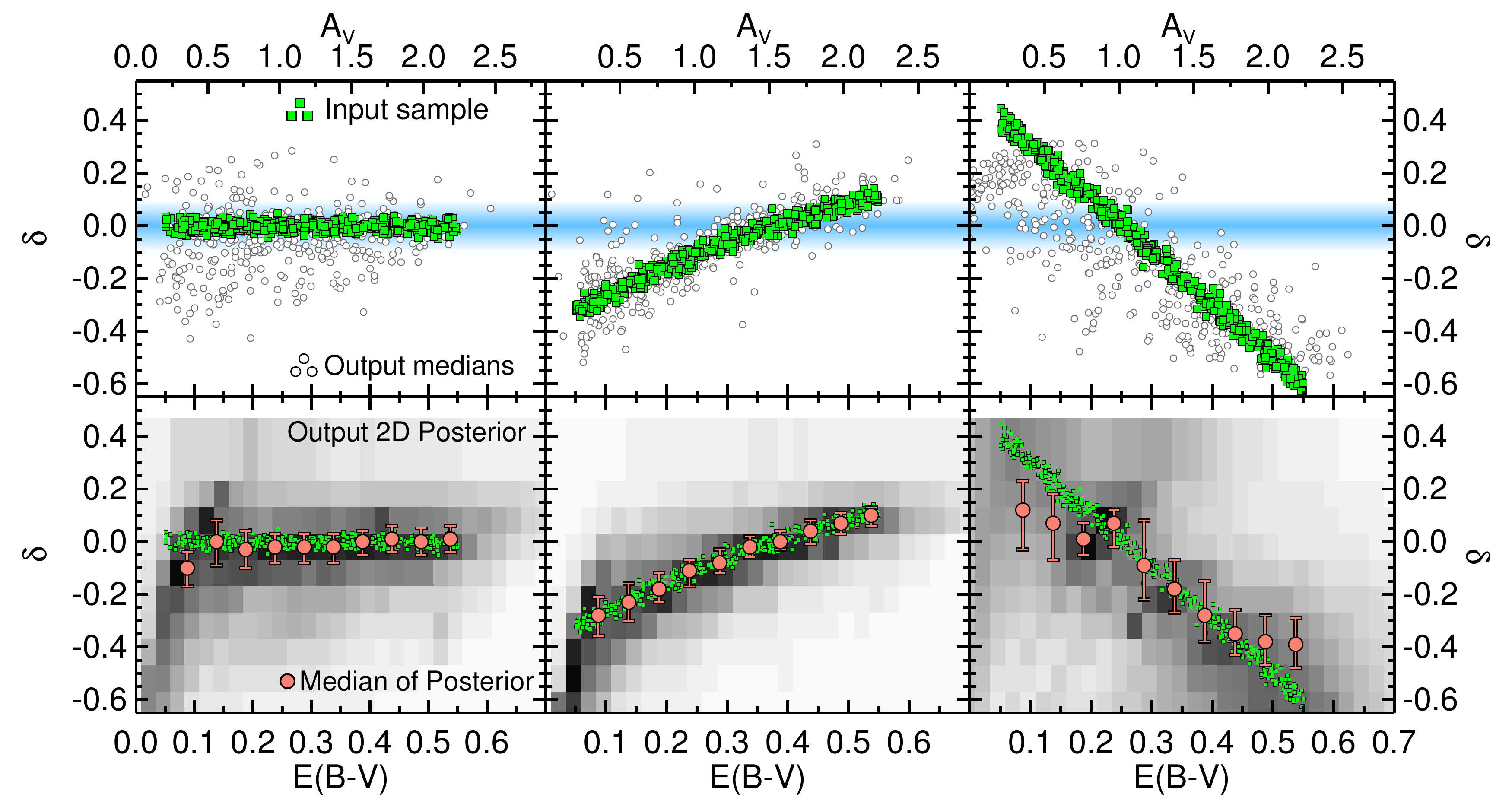}}
\caption{
\referee{
The tilt of the dust-attenuation curve, $\delta$, as a function of
\ebv\ for several input test samples to examine the robustness of the
main relation in \autoref{fig:photzDelta}. Green squares represent the
input \ebv\ and $\delta$ with a known model SED. The left panels show
a model where delta is constant with \ebv,  the right panels show a
model where delta declines with \ebv, and the middle panel shows a
model with the relation from \cite{Witt00} (shown in
\autoref{fig:Gordon}). The model fluxes are perturbed and assigned
uncertainty according to the real errors in the data. The recovered
median values of each $\delta$ and \ebv\ posterior is shown as grey
circles (top). The combined two-dimensional posterior for the whole
input sample is shown in the lower panels, where shaded regions
represent higher likelihood for the sample, salmon-colored circles
show the $\delta$ at median likelihood in bins of \ebv, and error bars
represent their 68~\% range in likelihood. 
}
}
\setcounter{figure}{15}
\end{figure*}

\referee{The results of the above test are unsurprising.  Redder
intrinsic SEDs will be fainter in the rest-UV, and by construction,
the fainter model fluxes are assigned higher uncertainties. Older ages
redden an SED in a similar manner as having high \ebv\ and high
$\delta$ or low \ebv\ and low $\delta$ especially when uncertainties
in the rest-UV are high.  This causes a broadening in the posteriors
of both $\delta$ and \ebv\ in a way that is correlated, causing a
covariance.} 

\referee{ Second, we then asked if the distribution of galaxy ages in
our sample is low enough such that the covariance does
not significantly bias the relation between \ebv\ and $\delta$.  We
addressed this with a similar test, where we fixed the age
distribution of galaxies to be the same as the one we measure for the
GOODS-S phot-$z$ sample.  The input \ebv\ was assigned to a random
value, and the input $\delta$ follows one of three test relationships:
$\delta$ is constant with \ebv,  $\delta$ increases with \ebv, and
$\delta$ decreases with \ebv.  For the increasing $\delta$ case, we
used the relationship from \cite{Witt00} (shown in
\autoref{fig:Gordon}) as the input. The input $\delta$ values were
also given a small random scatter to simulate a more realistic
relation and uncertainty in $\delta$.} 

\referee{We find that the input relation was well recovered for a
sample similar to the phot-$z$ sample of GOODS-S with a realistic
input age distribution.  \autoref{fig:AppendixDelta} summarizes the
results of this final test.  While the primary relation was recovered,
the covariances conspire to recover the increasing-$\delta$ relation
with less scatter and the decreasing-$\delta$ relation with more. In
addition, at \ebv $\lesssim 0.1$ the attenuation is not strong enough
to clearly distinguish between different dust laws. We conclude that
(1) we are able to recover several types of relationships on the plane
of \autoref{fig:photzDelta}, and (2) there may be some bias at \ebv
$\lesssim 0.1$, but because the attenuation is low regardless, the
effect is small. At \ebv =0.05, the uncertainty of
$\Delta(\delta)$=0.1 changes the attenuation at 1500\AA\ by only 10\%.
As supported by the observations in the $IRX-\beta$ relation, this
test gives credence that the relation in \autoref{fig:photzDelta} is
real and not an artifact of the SED-fitting procedure.}

\section{B. Using Herschel to Calculate the Total Infrared
Luminosities}\label{sec:AppendixLIR} The determinations of \lir\ used
in this work come from conversions of 24~\micron\ luminosity
calibrated by R13.  However, $\approx$~40\% of our sample have
detections at longer wavelengths with \herschel, providing the
opportunity to internally test our \lir\ measurements.  We determined
\lir\ by fitting several different suites of FIR SED templates to the
observed MIPS 24~\micron and/or \herschel\ PACS and SPIRE detections
of galaxies in our spec-$z$ sample. 

\autoref{fig:LIRcompare} shows several \lir\ calculations compared to
our fiducial \lir\ determined from the 24~\micron\ luminosity. We
compared our adopted \lir\ with \lir\ values derived from fitting the
full IR SED to \cite{Dale02} and \cite{Rieke09} templates, as well as
fitting to fluxes at 24$-$100~\micron\ only. In addition, we compare
the 24~\micron\ conversion to \lir\ proposed by \referee{
\cite{Wuyts08,Wuyts11b,Wuyts11a}. The R13 calibration is very similar
to that by Wuyts et al.  calibrations.  In all cases, the scatter in
the derived \lir\ is within $\sigma_\text{NMAD}~\approx~0.2$ dex.
This scatter is smaller than the correlations in $IRX-\beta$ found in
our primary results, and therefore our approximation of 24~\micron\
luminosity to \lir\ is reasonable. }

\referee{One benefit to using the 24~\micron\ flux to estimate \lir\
is that longer wavelength data, such as $>70-160$ \micron, could be
affected more by the the heating of cold dust from old stars.  Other
studies have shown that the longer wavelength data show more scatter
in the SFR calibration for this reason
\citep[see][]{Rieke09,Kennicutt09}.  Regardless, this is likely not a
significant factor in our sample (for example, dust heating from old stellar 
populations, which can lead to increased scatter, is mostly important 
for lower-luminosity galaxies, and accounts for
$<10\%$ of the total IR luminosity for \lir\ $> 10^{11}$ \Lsun\
\citep{Bell02, Calzetti10})} \referee{For these reasons the
24~\micron\ calibration of R13 (and others in the literature) should
be valid with a scatter of $\sim$0.2 dex. Lastly, we point out that
any uncertainty from the \lir\ calibration affects only the discussion
of the $IRX-\beta$ relation and not the derivation of the relation
between \ebv\ and $\delta$, which are based on the modeling of the
UV-to-near-IR SED and is independent of the total IR luminosities.}

\section{\referee{C. Changing the UV Slope, $\beta$}}\label{sec:AppendixBeta}
\referee{As described in \autoref{sec:Beta}, we ran tests to recover
the true $\beta$ when calculating $\beta$ from a power-law fit to the
UV or when using the best-fit SED. For our data, the
best-fit SED technique of \cite{Finkelstein12a} reproduced the
input $\beta$, while the power-law method produced some systematics
that worsened at higher redshifts. }
\referee{\autoref{fig:NewBeta} shows how the results of
\autoref{fig:money} and \autoref{fig:photzmoney} would change when
$\beta$ is calculated from a power-law fit to the observed UV bands.
For both the spec-$z$ and phot-$z$ samples, the position of galaxies
on $IRX-\beta$ broadly agrees with the favored dust law according to
the Bayes-factor evidence, in some cases more than the fiducial plots.
Importantly, small changes in $\beta$ do not change the clear divide
in $IRX$ between galaxies with opposing levels of Bayes-factor
evidence. For example, galaxies with evidence promoting a
starburst-like dust law tend to have higher $IRX$ than those promoting
an SMC-like dust law, at any given $\beta$. This provides evidence
that the method is correctly identifying galaxies with opposing dust
laws. }

\begin{figure}[!t]  \label{fig:LIRcompare}
\centerline{\includegraphics[scale=0.38]{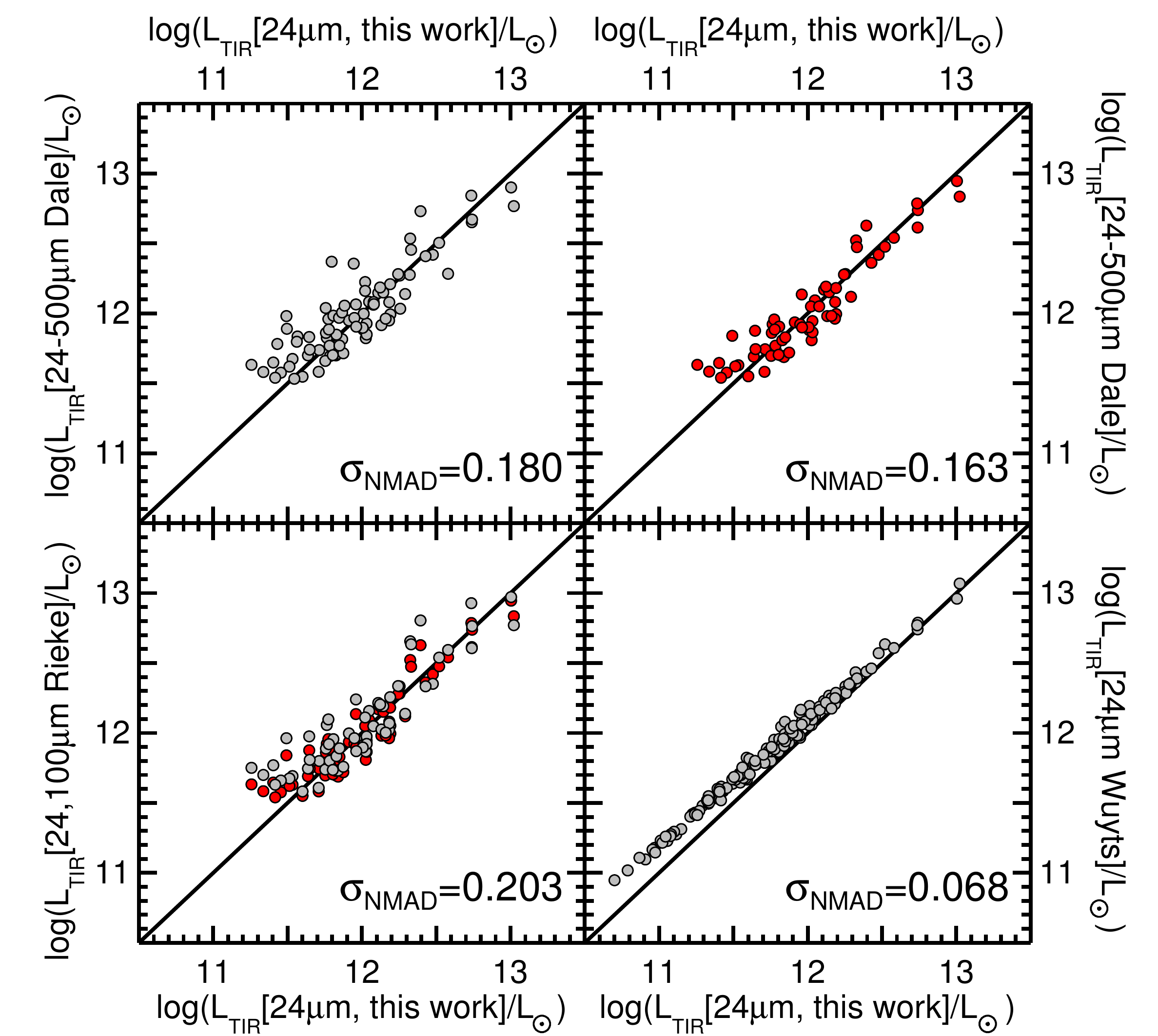}} 
\caption{
Galaxies in the spec-$z$ sample with \herschel\ PACS and/or SPIRE
detections are shown with several determinations of \lir\ as a
function of the fiducial 24~\micron\ method. The top panels assume
\cite{Dale02} fitting templates for all available FIR detections (top
left) or 24 and 100~\micron\ only (top right). The bottom left panel
assumes \cite{Rieke09} templates for 24 and 100~\micron\ bands (grey),
as well the Dale et al. assumption (red). The bottom right panel uses
the 24~\micron\ conversion to \lir\ from \cite{Wuyts08}. 
} 
\vspace{-0.2 cm}
\setcounter{figure}{16}
\end{figure}

\begin{figure}[!t]  \label{fig:NewBeta}
\centerline{\includegraphics[scale=0.57]{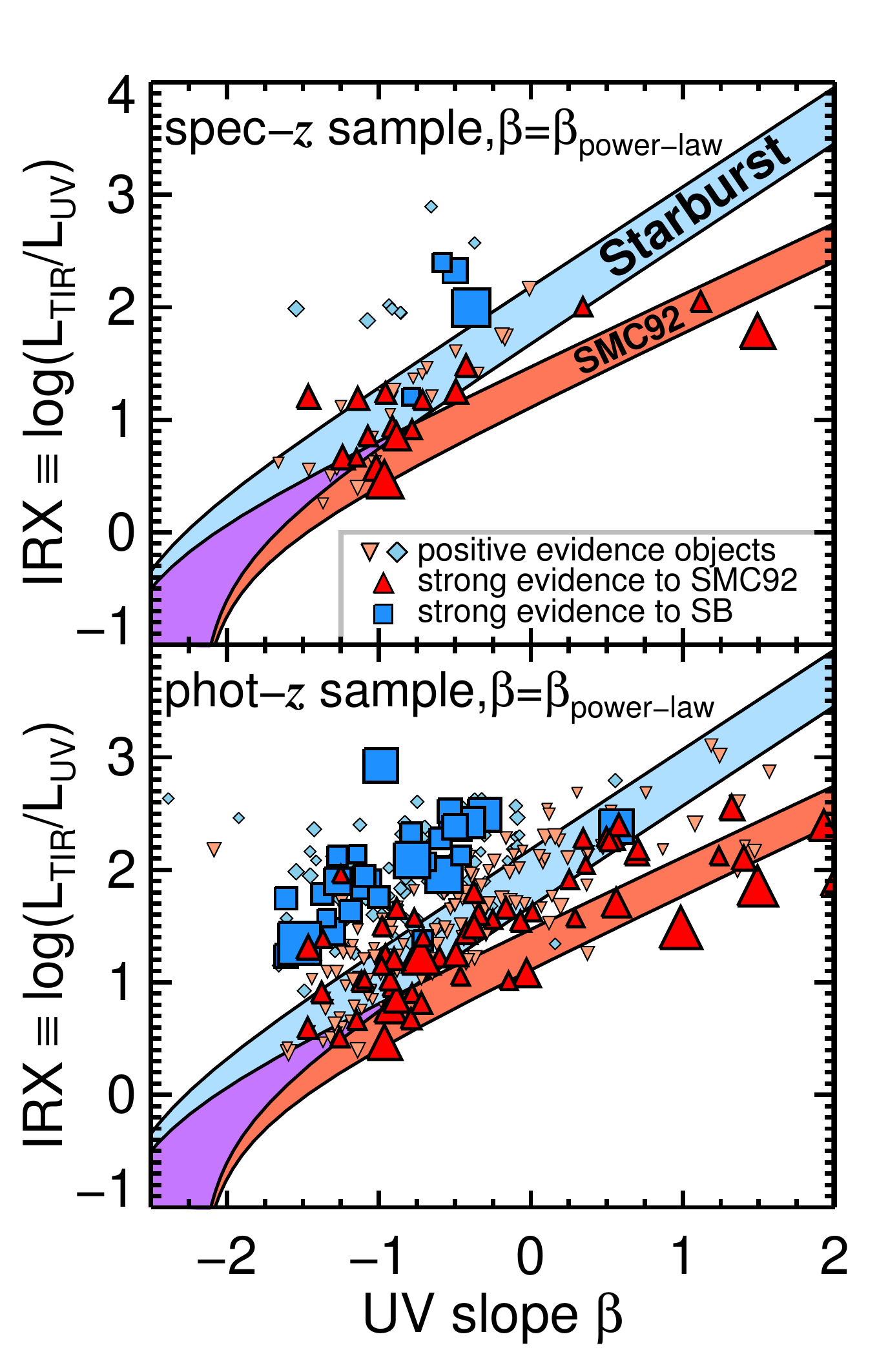}} 
\caption{
\referee{
The $IRX-\beta$ relation for the spec-$z$ (top) and phot-$z$ (bottom)
samples, where $\beta$ was derived from a power-law fit to the
observed rest-UV fluxes.
} 
} 
\setcounter{figure}{17}
\end{figure}

\begin{figure*}[!t]  
\centerline{\includegraphics[scale=0.53]{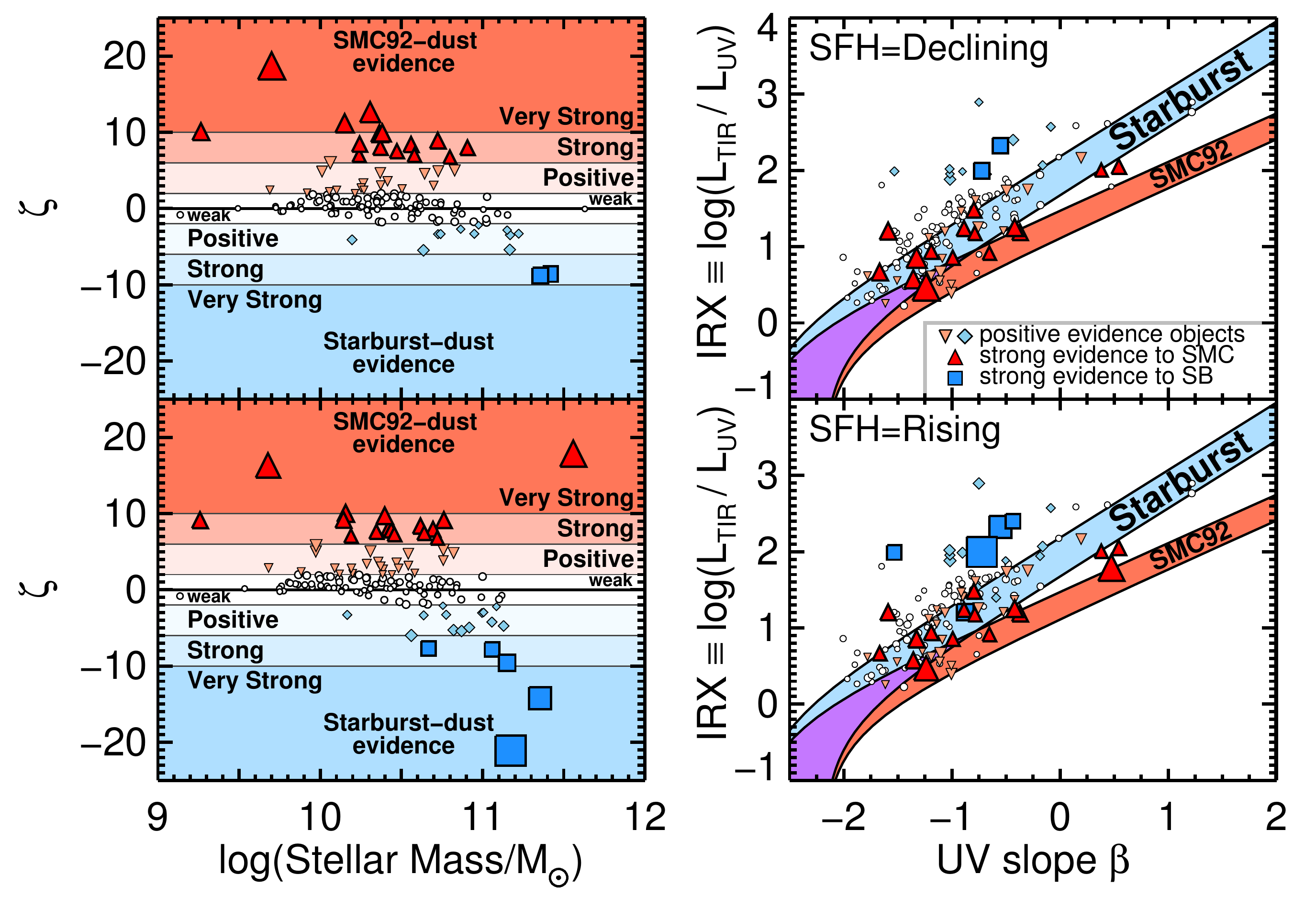}} 
\caption{                                                                      
Same as Fig.~\ref{fig:money}, but allowing the star formation history
to vary as fitted parameter for star formation rates that decline
(top) and rise (bottom) with time. Several colored tiers indicate the
selection of galaxies to have strong Bayes-factor evidence towards an
SMC92 (red triangles) or starburst (blue squares) dust law.  Although
the strength of evidence shifts when allowing the SFH to be free,
galaxies only ever lose or gain favorance towards a single dust law;
no galaxy changes the direction of its evidence. 
} 
\label{fig:IRXBeta_SFH}
\setcounter{figure}{18}
\end{figure*}

\section{D. Changing the Assumed Star-Formation \referee{History}} \label{sec:AppendixSFH}
For the primary results of this work, we used stellar population
models that assume a constant star-formation history. However, we must
consider if our choice of parameter space is missing models that could
mimic the SED-fitting evidence towards certain dust laws. In this
appendix, we allow the $e$-folding timescale ($\tau$), the time
interval over which the SFR is increased by a factor of $e$, to vary
as a parameter. We consider SFRs that rise and decline with cosmic
time with ranges described in Table~1. We also include the fits to
rising histories in \autoref{fig:ExampleSEDCALZ} and
\autoref{fig:ExampleSEDSMC} to illustrate that the additional
parameter does not create SED shapes \referee{that} mimic the evidence
towards different dust laws.

\autoref{fig:IRXBeta_SFH} shows how fitting the spec-$z$ sample with a
range of star-formation histories affects the selection of galaxies
based \referee{on} their Bayes-factor evidence and how that selection
propagates to the $IRX-\beta$ plane.  While the distribution of
Bayes-evidence shifts under the influence of a new parameter, there
are still galaxies with convincing evidence between dust laws. In
addition, no galaxy switches evidence from favoring the SMC92-like
dust to favoring starburst-like dust or vice versa. We conclude that
the evidence between dust laws is not an artifact of different
star-formation histories, at least for simple rising and declining
$\tau$-models.

\section*{Acknowledgements}
We thank the referee for thoughtful and constructive feedback on this
work. We acknowledge our colleagues in the CANDELS collaboration for
very useful comments and suggestions.  We also thank the great effort
of all the CANDELS team members for their work to provide a robust and
valuable data set.  We also thank Karl Gordon for insightful
discussions on the physical implications of these results.  This work
is based in part on observations taken by the CANDELS Multi-Cycle
Treasury Program with the NASA/ESA \emph{HST}, which is operated by
the Association of Universities for Research in Astronomy, Inc., under
NASA contract NAS5-26555. This work is supported by \emph{HST} program
No. GO-12060. Support for Program No.  GO-12060 was provided by NASA
through a grant from the Space Telescope Science Institute, which is
operated by the Association of Universities for Research in Astronomy,
Incorporated, under NASA contract NAS5-26555.  We acknowledge the
Spanish MINECO grant AYA2012-31277 for funding the contribution from
Pablo-Perez Gonzalez.  This work is based in part on observations made
with the \emph{Spitzer Space Telescope}, which is operated by the Jet
Propulsion Laboratory, California Institute of Technology under
contract with the National Aeronautics and Space Administration
(NASA).  The authors acknowledge the Texas A\&M University Brazos HPC
cluster that contributed to the research reported here.  URL:
http://brazos.tamu.edu.

\bibliographystyle{apj} \bibliography{salmon16}

\end{document}